\documentclass[11pt]{article}
\usepackage{amsfonts}
\usepackage{latexsym,amsmath}
\usepackage{amssymb,array}
\usepackage{calc}
\RequirePackage[dvips]{graphicx}
\usepackage{graphics}
\usepackage{longtable}
\usepackage{pdflscape}
\usepackage{authblk}
\usepackage{algpseudocode}
\usepackage{float}
\usepackage[left]{lineno}
\usepackage[colorlinks=true]{hyperref}
\hypersetup{urlcolor=blue, citecolor=red}
\makeatletter \oddsidemargin 0in \evensidemargin 0in \textwidth
16cm\textheight 20cm 
\begin{document}

\title{Control Charts for Percentiles of Truncated Beta Distributed Environmental Data Using Studentized Bootstrap Method}

\author[1]{Bidhan Modok}
\author[1]{Amarjit Kundu \footnote{Corresponding author e-mail:bapai\_k\ @yahoo.com}}
\author[2]{Shovan Chowdhury}
\affil[1] {Department of Mathematics, Raiganj University, West Bengal, India}
\affil[2] {Quantitative Methods and Operations Management Area, Indian Institute of Management Kozhikode, Kozhikode, India}

\maketitle

\begin{abstract}
This paper proposes a control chart for monitoring percentiles of a process that follows a truncated beta distribution, utilizing a studentized parametric bootstrap method to account for the case when in-control parameters are unknown. To evaluate the in-control performance, extensive Monte Carlo simulations are conducted across various combinations of percentiles, false alarm rates, and sample sizes, with performance measured in terms of the average run length. The out-of-control performance is thoroughly assessed by introducing shifts in the distributional parameters and comparing the proposed chart with the conventional beta-based chart. The effectiveness and practical applicability of the proposed chart is illustrated through real-world examples from environmental data.
\end{abstract}

\setcounter{section}{0}
\section{Introduction}
\setcounter{equation}{0}
\hspace*{0.2in} Control charts are widely recognized as the most effective graphical and statistical methods for monitoring and controlling process quality. While extensively used in the manufacturing sector, control charts also have broad applications in other domains such as healthcare (e.g., disease surveillance, patient wait times), business (e.g., monitoring service quality or stock price trends), software development (e.g., tracking code defects per release or website load times), and environmental studies (e.g., observing pollutant levels or natural resource usage). The primary purpose of a control chart is to monitor the stability of key process parameters, such as the mean, variance, or proportion, and to detect and eliminate sources of variation. \\
\hspace*{0.2in} Traditionally, $p$ and $np$ control charts, originally proposed by Shewhart (1924), have been widely employed to monitor the proportion of nonconforming items as a measure of attribute performance. These charts are constructed under the assumption that the underlying data follow a binomial distribution. For large sample sizes, control limits are typically derived using a normal approximation to the binomial distribution (Sant'Anna and ten Caten, 2012). However, when the sample size is large and the proportion of nonconforming items is small (e.g., p = 0.005, 0.05, 0.1), the binomial distribution can be approximated by a Poisson distribution, which is notably skewed. In such cases, the resulting normal approximation becomes unreliable. This issue is particularly pronounced in high-quality processes, where the defect rates are low, rendering conventional $p$ and $np$ charts less effective (Wang, 2009; Bersimis \emph{et al.}, 2014). Additionally, when the sample size is not sufficiently large, the assumption of normality becomes untenable, further compromising the chart’s ability to accurately monitor process performance (Wetherill and Brown, 1991; Xie and Goh, 1993). Another known drawback is that the computed control limits may fall outside the (0, 1) range, which lacks physical interpretability (Bersimis \emph{et al.}, 2014). Several modifications to traditional proportion charts have been proposed to address these limitations (Quesenberry, 1991; Heimann, 1996; Schwertman and Ryan, 1997; Chen, 1998). Nonetheless, it is important to recognize that proportion data are not always the result of Bernoulli trials. In many applications, proportions arise from ratios of continuous quantities, such as traffic intensity (arrival rate to service rate), chemical composition in pharmaceuticals, alcohol content in beverages, alloy proportions in manufacturing, and unemployment rates. Proportion data are also prevalent in hydrological and environmental studies (Bayer and Cribari-Neto, 2017), such as relative humidity (RH) which is defined as the ratio of the partial pressure of water vapor to the equilibrium vapor pressure of water. \\
\hspace*{0.2in} Given these considerations, it is evident that standard $p$ and $np$ charts, and even their modified versions may be inadequate for reliably monitoring proportion-type data across a wide range of practical contexts. The beta distribution, defined on the interval [0, 1], has long been a popular and flexible choice for modeling proportion data due to its diverse range of shapes and wide applicability (Gupta and Nadarajah, 2004; Johnson \emph{et al.}, 1995). Recognizing this suitability, Sant'Anna and ten Caten (2012) proposed a control chart that assumes a beta distribution to model proportion data. Building upon this framework, Bayer \emph{et al.} (2018) introduced a beta regression control chart, which generalizes the earlier model by allowing the beta-distributed quality characteristic to be related to control variables (covariates). Further developments in this area include the work of Ho \emph{et al.} (2019), who evaluated the performance of beta control charts with control limits derived from simplex and unit gamma distributions, assuming known parameters of the beta distribution. Additionally, de Araujo \emph{et al.} (2019) proposed a beta control chart designed specifically to monitor zero-inflated processes, addressing scenarios where a substantial proportion of observations are exactly zero, while Lima \emph{et al.} (2023) addressed a similar problem with zero or one-inflated data. Later, de Araujo \emph{et al.} (2021) and Silva \emph{et al.} (2025) proposed control charts for proportion assuming Kumaraswamy and unit gamma distribution, respectively. Chowdhury et al. (2022) first argued in favor of monitoring percentiles of proportion data coming from a beta-distributed process.\\
\hspace*{0.2in} In many practical scenarios, processes do not permit extremely high proportions of nonconforming items (or extremely low proportions of conforming ones). In such cases, right or left truncation of the beta distribution can more accurately reflect the underlying randomness in the proportion data compared to the standard, untruncated beta distribution. From this perspective, employing a distribution truncated over a relevant subinterval of the [0, 1] range may provide a more appropriate modeling framework, particularly for highly reliable processes, where the observed proportions lie within a narrower range. Ignoring truncation can lead to biased estimates of both the mean and variance of the proportion variable. The motivation for this study arises from the observation that, in many applications, proportion data are effectively bounded within a limited subrange of [0, 1], rather than the full interval. Another practical example where truncating the beta distribution is appropriate arises in the monitoring of relative humidity, which typically assumes values within a restricted subinterval of [0, 1], depending on the seasonal month or geographic location under consideration. Consequently, the truncated beta (Tbeta) distribution is proposed as a more suitable alternative to model such data. This rationale supports the use of the Tbeta distribution as the basis for constructing an effective control chart for monitoring proportion data.\\ 
\hspace*{0.2in} The maximum likelihood estimators of the parameters of the Tbeta distribution do not have closed-form expressions. As a result, the sampling distribution of Tbeta percentiles cannot be derived analytically. This limitation necessitates the use of computational techniques, such as the parametric bootstrap method, to construct the control limits for the proposed Tbeta control chart (hereafter referred to as the Tbeta chart). In particular, prior knowledge of the skewed nature of the data relevant to the application motivates the use of a parametric bootstrap to enhance accuracy. Efron (1979) distinguishes between two bootstrap procedures: the nonparametric bootstrap, which relies on the empirical distribution generated from a sample of size n drawn from an unknown distribution $F$, and the parametric bootstrap, which assumes $F$ belongs to a known parametric family. Given the distributional assumptions in this study, the parametric studentized bootstrap approach is adopted to estimate control limits effectively. For further discussions on these techniques, readers are referred to Efron and Tibshirani (1993), Liu and Tang (1996), Seppälä \emph{et al.} (1996), and Jones and Woodall (1998). Parametric bootstrap-based control charts for quantiles have gained considerable attention in the statistical process control (SPC) literature (see Nichols and Padgett, 2005; Lio and Park, 2008, 2010; Lio \emph{et al.}, 2014; Chiang \emph{et al.}, 2017, 2018; Chowdhury \emph{et al.}, 2025; Modok \emph{et al.}, 2025). To the best of our knowledge, no study has explored the application of the bootstrap approach to percentiles from the Tbeta distribution for monitoring proportion data. This gap in the literature motivates the present work, where we construct a bootstrap-based control chart specifically for monitoring percentiles of RH data following a Tbeta distribution. \\
\hspace*{0.2in} The remainder of the paper is structured as follows. Section 2 introduces the proposed Tbeta control chart along with the underlying statistical framework. Section 3 focuses on the practical implementation of the chart, including the tabulation of control limits and average run length ($ARL$). This section also presents the in-control (IC) and out-of-control (OOC) performance of the Tbeta chart, evaluated through an extensive Monte Carlo simulation, and includes a comparative analysis with the conventional beta chart. Section 4 illustrates two application of the Tbeta chart from environmental process. Finally, Section 5 concludes the paper.
\setcounter{section}{1}
\section{Construction of Studentized Bootstrap Tbeta Control Chart}
\setcounter{equation}{0}
In this section, the methodology of the proposed control chart is discussed.
\subsection{Statistical Framework}
Let $X$ be a random variable following Tbeta distribution with parameters $\theta_1$ and $\theta_2$ with $a$ and $b$ as lower and upper truncation points respectively. The probability density function (pdf) and cumulative distribution function (cdf) of $X$ are given by  
\begin{equation}\label{e1}
f(x|\theta_1,\theta_2)=\frac{1}{I_b(\theta_1,\theta_2)-I_a(\theta_1,\theta_2)} x^{\theta_1-1}(1-x)^{\theta_2-1} , 0\leq a\leq x\leq b\leq 1 ,\theta_1>0,\theta_2>0,
\end{equation}\
and 

\begin{equation}\label{e2}
F(x|\theta_1,\theta_2)=
    \begin{cases}
        0, & \text{for } x < a,\\
      \frac{\int_0^xu^{\theta_1-1}(1-u)^{\theta_2-1}du}{I_b(\theta_1,\theta_2)-I_a(\theta_1,\theta_2)}, & \text{for } a\leq x\leq b,\\
			1, &\text{for } x > b.
    \end{cases}
\end{equation}

where $$ I_c(\theta_1,\theta_2)=\int_0^cu^{\theta_1-1}(1-u)^{\theta_2-1}du.$$
The Tbeta distribution with $a=0$ and $b=0.5$ is found to exhibit both positive and negative skewness for various choices of $\theta_1$ and $\theta_2$ as shown in the following Figure $\ref{figure4}$.
\begin{figure}[ht]
\centering
\begin{minipage}[b]{0.48\linewidth}
\includegraphics[height=6.8 cm]{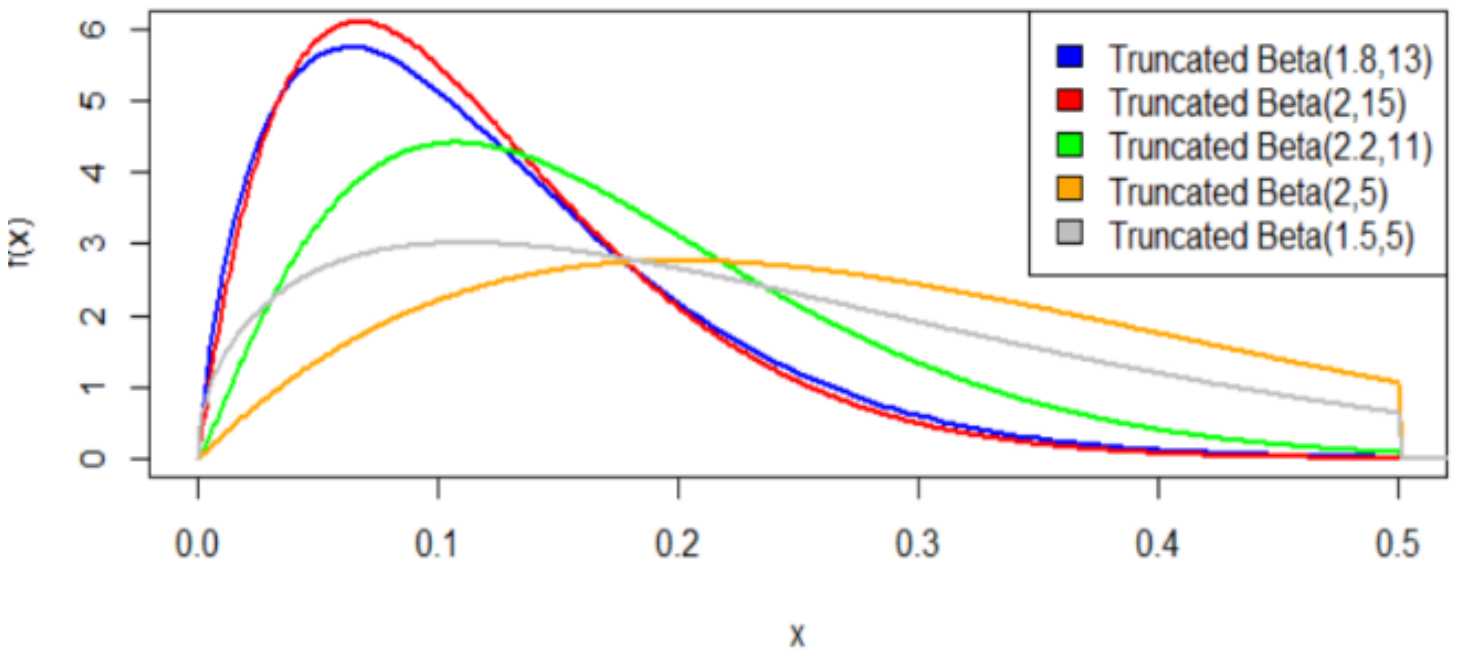}
\end{minipage}\label{figure3}
\quad
\begin{minipage}[b]{0.48\linewidth}
\includegraphics[height=6.8 cm]{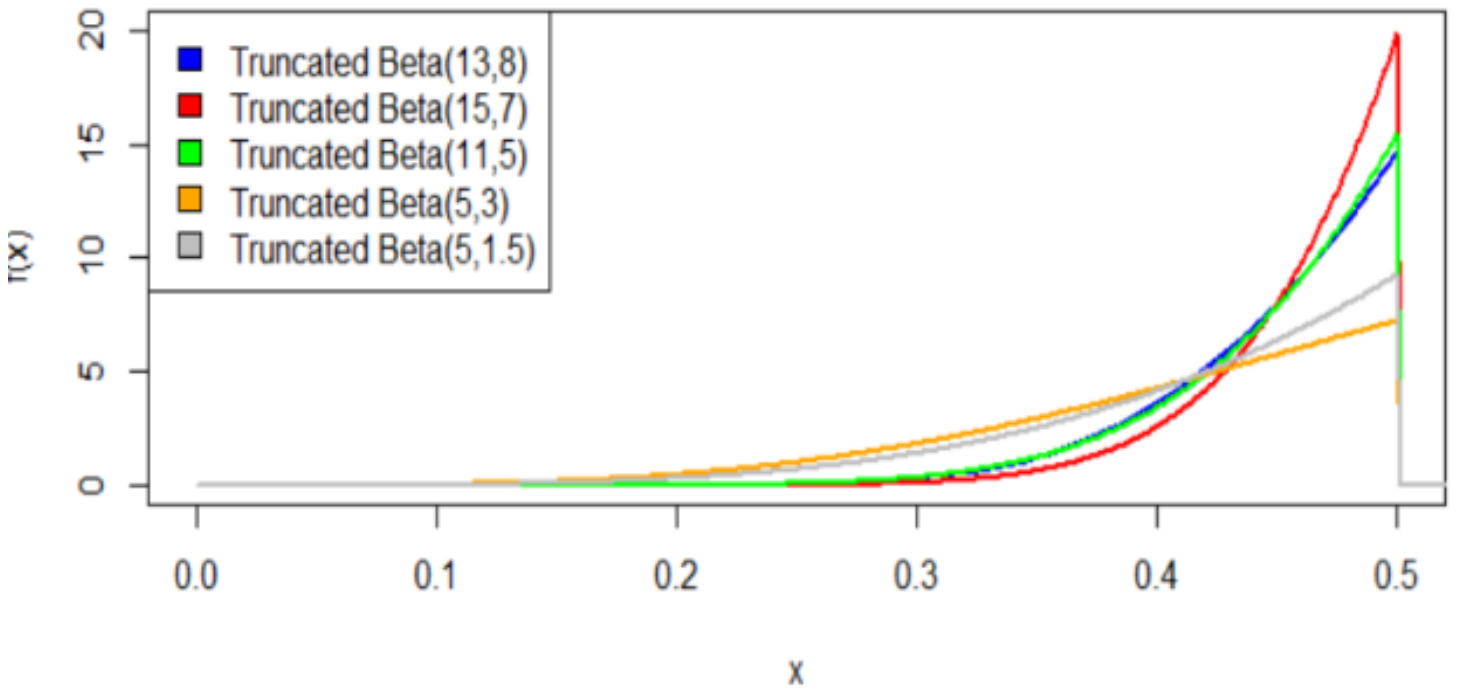}
\end{minipage}\caption{\label{figure4} Tbeta distribution for various choices of $\theta_1$ and $\theta_2$}
\end{figure}
Now, let $X_i=\left(x_{i1},x_{i2},\ldots,x_{in}\right)$ be $i^{th}$ in-control (IC) random subgroup of size $n~(i =1, 2,\ldots, k)$ drawn from phase I process control procedure following Tbeta distribution as given in $(\ref{e1})$. The log-likelihood function of $k$ such independent random subgroups (sample) of size $n$ each is given by
\begin{equation}\label{e3}
\mathrm{ln}L(\theta_1,\theta_2)=\sum_{i=1}^k\sum_{j=1}^n\mathrm{ln}f(x_{ij}|\theta_1,\theta_2).
\end{equation}
The MLEs $\theta_1$ and $\theta_2$, are obtained by maximizing the loglikelihood function as shown in (\ref{e3}), and subsequently solving the non-linear equations $\frac{\partial \mathrm{ln}L}{\partial \theta_1}=0$ and $\frac{\partial \mathrm{ln}L}{\partial \theta_2}=0$. As closed-form solutions of the MLEs are not available, the MLEs are obtained by using numerical optimization methods in R (version $4.0.2$). The detailed computational framework is mentioned in the next subsection.
Let $\xi_p$ denote the $100 p^{th}$ percentile of Tbeta distribution obtained from the following equation
\begin{equation}\label{e4}
\xi_p=F^{-1}(p;\theta_1,\theta_2).
\end{equation}
Therefore, the MLE of the $100p^{th}$ percentiles, denoted by $\hat\xi_p$, is given by 
\begin{equation}\label{e5}
\hat{\xi_p}=F^{-1}(p;\hat{\theta_1},\hat{\theta_2}).
\end{equation}
The proposed chart is developed in the next subsection.
\subsection{Proposed Charting Procedure}
This section outlines the step-by-step procedure for constructing studentized parametric bootstrap process control chart to monitor the quantiles of data following TBeta distribution. Hereafter, the term $SBTBC$ is used to refer to the Studentized Bootstrap TBeta Control Chart.
\begin{enumerate}
	\item[{\bf Step-1:}] Collect and establish $k$ reference samples $X^i_n=(x_{i1},x_{i2},\ldots,x_{in})$, $i=1,2,\ldots, k$, of size $n$ each from an IC process (Phase I process) following cdf $F(x|\theta_1,\theta_2 )$ as in $(\ref{e2})$.
	\item[{\bf Step-2:}] Obtain the MLEs of  $\theta_1$ and $\theta_2$ from the collection of $(n\times k=m)$ sampling units from Step-$1$ and using $(\ref{e5})$  estimate $\hat{\xi}_p$.
	\item[{\bf Step-3:}] Generate a bootstrap sample of size $n$, $x_1^*,x_2^*,\ldots,x_n^*,$ from the $m$ sampling units as obtained in Step-$1$.
	\item[{\bf Step-4:}] Obtain the MLEs of $\theta_1$ and $\theta_2$  using the bootstrap sample obtained in Step-$3,$ and denote these as $\hat\theta_1^*$ and $\hat\theta_2^*.$
	\item[{\bf Step-5:}] Using $(\ref{e5})$ compute the bootstrap estimate of the $p^{th}$ quantile as
	\begin{equation} \label{e13}
		\hat\xi_p^*=F^{-1}(p; \hat\theta_1^*,\hat\theta_2^*).
	\end{equation}
	\item[{\bf Step-6:}] Repeat Steps $3$-$5$ large number of times ($B$, say) to obtain bootstrap estimates of $\hat\xi_p^*$, denoted by $\hat\xi_{1p}^*,\hat\xi_{2p}^*,\ldots,\hat\xi_{Bp}^*.$
	\item[{\bf Step-7:}] Compute the mean and standard error of $\hat\xi_p^*$ as $\overline{\xi}_p^*$=$\frac{1}{B} \sum_{i=1}^B \hat\xi_{ip}^*$ and $SE(\hat\xi_p^*)$=$\sqrt{\frac{1}{B}\sum_{i=1}^B\left(\hat\xi_{ip}^*-\overline{\xi}_p^*\right)^2}$ respectively.
	\item[{\bf Step-8:}] Obtain studentized quantile as $t^*_ {ip}$=$\frac{\hat\xi_{ip}^*-\hat{\xi}_{p}}{SE(\hat\xi_p^*)},~i=1,2,...,B,$ where $\hat{\xi}_{p}$ is obtained from Step-$2$.
	\item[{\bf Step-9:}] Using $B$ studentized quantiles from Step $8$, obtain the $\frac{\nu}{2}^{th}$ and $(1- \frac{\nu}{2})^{th}$ studentized quantiles empirically and refer them as $t^*_{Lp}$ and $t^*_{Up}$ respectively. By reversing the studentized quantile, the lower and upper control limits of the chart can be obtained as $LCL=\overline{\xi}_p^*+t^*_{\xi_ {Lp}} \times SE(\hat\xi_p^*)$ and $UCL=\overline{\xi}_p^*+t^*_{\xi_{Up}} \times SE(\hat\xi_p^*)$ respectively. Here, $\nu,$ the false alarm rate ($FAR$) is defined as the probability that an observation is considered out of control (OOC) when the process is actually in IC. 
	\item[{\bf Step-10:}] Sequentially observe the $j^{th}$ phase II (test) sample $Y_{j:n}=(Y_{j1},Y_{j2},\ldots,Y_{jn})$ of size $n, j=1,2,…$.
	\item[{\bf Step-11:}] Sequentially obtain $\hat{\xi}_{jp}$ using (\ref{e13}) after obtaining MLEs of the parameters using the $j^{th}$ test sample as described in Step-$2.$   
	\item[{\bf Step-12:}] Plot $\hat{\xi}_{jp}$ against $LCL$ and $UCL$ as obtained in Step-$9$ of the Phase I process.
	\item[{\bf Step-13:}] If $\hat{\xi}_{jp}$ falls between the $LCL$ and $UCL,$ then the process is assumed to be in-control, otherwise, an OOC signal is activated.
\end{enumerate}

\setcounter{section}{2}
\section{Simulation and Performance Analysis}
\setcounter{equation}{0} 
This section presents a comprehensive simulation study to evaluate the in-control (IC) and out-of-control (OOC) performance of the $SBTBC$ for monitoring percentiles. Numerical computations are performed in $R$ (version $4.3.2$) using Monte Carlo simulations to estimate the average $UCL$ and $LCL$. The MLEs of the TBeta distribution parameters $\theta_1$ and $\theta_2$ are obtained for the true choice of $\theta_1 = 2$, and $\theta_2 = 15$, representing a right-skewed distribution under different truncation intervals ($a=0,\;b=0.5;\, a=0, \, b=0.6;\, a=0.1, \, b=0.6$). Control limits are calculated based on $B = 5000$ bootstrap samples. Simulations are conducted using sample size $n = 10$, subgroup size $k = 20$, various percentiles $(p=0.1,0.25,0.5,0.75,0.9)$, and FARs $(\nu = 0.005, 0.0027, 0.002).$ Chart performance is evaluated using run length (the number of samples before the first OOC signal), with average run length ($ARL$) and standard deviation of run length ($SDRL$) calculated from $5,000$ replications.

\subsection{IC Performance}
\hspace*{0.2in} The estimated in-control (IC) control limits for the $SBTBC$ chart are presented in Table $\ref{tab2}$ for three different truncation intervals: i) $a=0$, $b=0.5,$ ii)  $a=0$, $b=0.6,$ and iii) $a=0.1$, $b=0.6.$ Along with the control limits, the table includes key performance measures $ARL$ and $SDRL$, denoted by $ARL_0$, and $SDRL_0$. Theoretically, the reciprocal of $FAR$ corresponds to the nominal $ARL.$ Thus, for $\nu=0.005,0.0027,0.002$, the expected nominal $ARL$s are $200$, $370,$ and $500$ respectively. In general, lower $ARL$ values imply narrower control limits, while higher $ARL$ values indicate wider limits, resulting in fewer false alarms. The simulated $ARL_0$ values closely match these theoretical expectations, aligning with the geometric distribution model of run length with mean $\frac{1}{\nu}$. This confirms that the $SBTBC$ chart performs effectively across different levels of data skewness. Furthermore, the observed $SDRL_0$ values are consistent with the theoretical behavior of the geometric distribution, further validating the robustness of the proposed chart under IC conditions.

\begin{center}
	\tiny{\begin{longtable}{|c|c|c|c|c|}
			\caption{ Control limits, $ARL_0$ and $SDRL_0$ for $\theta_1=2$, $\theta_2=15$  and $n=10$ in $SBTBC$ scheme}\label{tab2}\\
			\hline
			\endfirsthead
			\multicolumn{5}{l}%
			{\tablename\ \thetable\ -- \textit{Continued from previous page}} \\
			\hline
			\endhead
			\hline \multicolumn{5}{r}{\textit{Continued on next page}} \\
			\endfoot
			\hline
			\endlastfoot
			$\nu$ & $LCL$  & $UCL$ &  $ARL_0$ & $SDRL_0$\\
			\hline
			\multicolumn{5}{|c|}{$a=0$, $b=0.5$} \\ \hline
			
			\multicolumn{5}{|c|}{$p=0.1$}\\ \hline
			0.005 & 0.0118 & 0.1128 & 190.3678 & 192.1649 \\ \hline
			0.0027 & 0.0066 & 0.1066 & 364.4794 & 362.6929 \\ \hline
			0.002 & 0.0044 & 0.1126 & 495.6441 & 502.9730 \\ \hline
			\multicolumn{5}{|c|}{$p=0.25$}\\ \hline
		0.005 & 0.0235 & 0.1390 & 193.3240 & 193.5018 \\ \hline
		0.0027 & 0.0231 & 0.1388 & 371.1598 & 364.9552 \\ \hline
		0.002 & 0.0245 & 0.1201 & 478.8390 & 480.3571 \\ \hline
		\multicolumn{5}{|c|}{$p=0.5$}\\ \hline
		0.005 & 0.0547 & 0.1936 & 200.4016 & 200.1543 \\ \hline
		0.0027 & 0.0430 & 0.1830 & 365.0660 & 372.9870 \\ \hline
		0.002 & 0.0519 & 0.2103 & 511.4801 & 510.1052 \\ \hline
		\multicolumn{5}{|c|}{$p=0.75$}\\ \hline
		0.005 & 0.0710 & 0.2638 & 201.5152 & 207.1191 \\ \hline
    	0.0027 & 0.0241 & 0.1305 & 354.5344 & 358.6505 \\\hline
		0.002 & 0.0248 & 0.1409 & 506.9778 & 510.3991 \\ \hline
			\multicolumn{5}{|c|}{$p=0.9$}\\ \hline
			0.005 & 0.0941 & 0.3791 & 205.6484 & 206.4490 \\ \hline
			0.0027 & 0.1051 & 0.3934 & 356.3640 & 356.8699 \\ \hline
			0.002 & 0.0840 & 0.3226 & 518.1251 & 521.6429 \\ \hline
			\multicolumn{5}{|c|}{$a=0$, $b=0.6$}\\
	       \hline
			\multicolumn{5}{|c|}{$p=0.1$}\\\hline
			0.005 & 0.0048 & 0.0919 & 203.7220 & 203.7070 \\ \hline
			0.0027 & 0.0073 & 0.0323 & 365.0071 & 368.8691 \\ \hline
			0.002 & 0.0023 & 0.0973 & 491.6991 & 492.8851 \\ \hline
			
			\multicolumn{5}{|c|}{$p=0.25$}\\\hline
			0.005 & 0.0175 & 0.1340 & 204.9231 & 199.0096 \\ \hline
			0.0027 & 0.0281 & 0.1379 & 397.0881 & 399.1530 \\ \hline
			0.002 & 0.0209 & 0.1420 & 490.8090 & 480.3295 \\ \hline
			
			\multicolumn{5}{|c|}{$p=0.5$}\\ \hline
			0.005 & 0.0484 & 0.1689 & 194.9581 & 195.5317 \\ \hline
			0.0027 & 0.0509 & 0.1911 & 365.5595 & 367.0453 \\ \hline
			0.002 & 0.0479 & 0.1957 & 481.4226 & 479.7578 \\ \hline
			
			\multicolumn{5}{|c|}{$p=0.75$}\\ \hline
			0.005 & 0.0737 & 0.2600 & 195.0973 & 199.7507 \\ \hline
			0.0027 & 0.0707 & 0.2595 & 356.9395 & 352.6593 \\ \hline
			0.002 & 0.0677 & 0.2340 & 501.0161 & 504.0759 \\ \hline
			
			\multicolumn{5}{|c|}{$p=0.9$}\\ \hline
			0.005 & 0.1009 & 0.3379 & 198.5759 & 199.7311 \\ \hline
			0.0027 & 0.0983 & 0.3430 & 353.0329 & 351.2478 \\ \hline
			0.002 & 0.0894 & 0.3824 & 513.8097 & 509.2138 \\ \hline
        	\multicolumn{5}{|c|}{$a=0.1$, $b=0.6$}\\
        	\hline
			\multicolumn{5}{|c|}{$p=0.1$}\\\hline
			0.005 & 0.1052 & 0.1684 & 196.0267 & 204.9942 \\ \hline
			0.0027 & 0.1048 & 0.1700 & 349.9626 & 347.0136 \\ \hline
			0.002 & 0.1037 & 0.1622 & 504.0588 & 512.0527 \\ \hline
			
			\multicolumn{5}{|c|}{$p=0.25$}\\\hline
			0.005 & 0.1113 & 0.1869 & 203.3101 & 207.7362 \\ \hline
			0.0027 & 0.1115 & 0.1899 & 358.9015 & 365.1276 \\ \hline
			0.002 & 0.1120 & 0.1947 & 483.2564 & 489.3757 \\ \hline
			
			\multicolumn{5}{|c|}{$p=0.5$}\\ \hline
			0.005 & 0.1231 & 0.2237 & 195.3392 & 197.1604 \\ \hline
			0.0027 & 0.1207 & 0.2308 & 369.5437 & 369.4181 \\ \hline
			0.002 & 0.2498 & 0.2498 & 493.3818 & 501.4982 \\ \hline
			
			\multicolumn{5}{|c|}{$p=0.75$}\\ \hline
			0.005 & 0.1392 & 0.2995 & 200.4383 & 208.5294 \\ \hline
			0.0027 & 0.1428 & 0.3178 & 364.7736 & 367.0545 \\ \hline
			0.002 & 0.1357 & 0.1375 & 522.5044 & 519.1247 \\ \hline
			
			\multicolumn{5}{|c|}{$p=0.9$}\\ \hline
			0.005 & 0.1527 & 0.3831 & 198.7895 & 201.0358 \\ \hline
			0.0027 & 0.1490 & 0.3939 & 356.3915 & 357.6487 \\ \hline
			0.002 & 0.1490 & 0.4066 & 514.7155 & 507.1302 \\ \hline
	
	\end{longtable}}
\end{center}

\subsection{OOC Performance}
\hspace*{0.2in} In order to investigate the OOC performance of the proposed $SBTBC$ chart, we measure the impact of changes in the IC parameter estimates on $ARL_1$. In other words, we consider sample taken from TBeta$(\theta_1+\Delta\theta_1,\theta_2+\Delta\theta_2)$, while the IC sample coming from TBeta$(\theta_1,\theta_2)$, and examine the effects of shifts $\Delta\theta_1$ and/or $\Delta\theta_2$ on the $ARL_1$ of the percentile of Tbeta distribution. For $\theta_1=2,\theta_2=15,\nu=0.0027$, these studies are done for two different cases: i) When $a=0$ and $b=0.5$ (see Table \ref{tab5}) and ii) when $a=0.1$ and $b=0.6$ (see Table \ref{tab6}). In general, the simulation results reveal that for fixed $n$, the OOC $ARL_1$ values, denoted as $ARL_1$ for the percentiles all decrease sharply with both downward and upward small, medium, and large shifts in the parameters. The phenomenon indicates that the $SBTBC$ chart is effective in detecting shifts in the parameters. However, the extent of effectiveness of the chart i.e, the speed of detection of shifts varies depending on the type of shift, the parameters, and the percentile being considered. It is observed from Table $\ref{tab5}$ and Figure $\ref{OOC}[(a),(b)]$ that the chart detects downward shift in both $\theta_1$ (when $\theta_2$ is IC ) and  $\theta_2$ faster ($\theta_1$ is IC)  than  their upward shift irrespective of the choice of percentile. For example it is observed that for a $4\%$ decrease in $\theta_1$ $(\theta_2)$ when $\theta_2$ $(\theta_1)$ is in IC , there is about $22.85\%\ (22.98\%)$ reduction in the $ARL_1$ of the median while there is about  $10.05\%\ (5.17\%)$ for $4\%$ increase in $\theta_1$ $(\theta_2)$ when $\theta_2$ $(\theta_1)$ is in IC.

Again, from Figure $\ref{OOC}$(d) it is clear that for $2\%$ deviation in $\theta_2$, the $ARL_1$s around $10^{th}$ and $25^{th}$ percentile are smaller that the other percentiles for both upward and downward shift of $\theta_1$. More  specifically The $ARL_1$ around $10^{th}~(25^{th})$ are found to be smaller than the other percentile for upward (downward) shift of $\theta_1$.
For example, $2\%$ increases in $\theta_2$ and $4\%$ increases in $\theta_1$, the $ARL$ around $10^{th}$ percentile is decreased in $17.02\%$ while  the $ARL_1$ around $25^{th}$,  $50^{th}$, $75^{th}$ and $90^{th}$ are decreased by $12.09\%$,$2.85\%$,$7.03\%$ and $10.61\%$ respectively.\\
 Also from Figure $\ref{OOC}$(c) it is clear that for $2\%$ deviation in $\theta_1$, the $ARL_1$s around $90^{th}~(75^{th})$ are found to be smaller than the other percentile for upward (downward) shift of $\theta_2$. For example, $2\%$ increases in $\theta_1$ and $4\%$ decreases in $\theta_2$, the $ARL$ around $75^{th}$ percentile is decreased in $39.79\%$ while  the $ARL_1$ around $10^{th}$,  $25^{th}$, $50^{th}$ and $90^{th}$ are decreased by $35.26\%$,$30.51\%$,$33.53\%$ and $39.08\%$ respectively.

\begin{center}
	{\tiny 
		\begin{longtable}{|c|c|c|c|c|c|}
			\caption{OOC performance of Tbeta Percentiles for $\theta_1=2, \theta_2=15, $a=0$, $b=0.5$,  \nu=0.0027$}\label{tab5}\\
			\hline
			 & $p=0.1$ & $p=0.25$ & $p=0.5$ & $0.75$ & $0.9$\\
			\hline
			\endfirsthead
			\multicolumn{6}{l}%
			{\tablename\ \thetable\ -- \textit{Continued from previous page}} \\
			\hline
			$\Delta\theta_2$ & $p=0.1$ & $p=0.25$ & $p=0.5$ & $0.75$ & $0.9$\\
			\hline
			\endhead
			\hline \multicolumn{6}{r}{\textit{Continued on next page}} \\
			\endfoot
			\hline
			\endlastfoot
			\multicolumn{6}{|c|}{$\Delta\theta_1=-0.02$}\\
			\hline
		$\Delta\theta_2$	& $ARL_1(SDRL_1)$ & $ARL_1(SDRL_1)$ & $ARL_1(SDRL_1)$ & $ARL_1(SDRL_1)$ & $ARL_1(SDRL_1)$\\
			\hline
			-0.3	&39.312(40.668)	&21.083(21.623)	&10.951(11.476)&	10.291(10.906)	&12.293(13.006)\\ \hline
			-0.2	& 98.276(97.314)& 67.222(67.587) &	41.492(41.012)&	36.322(36.091)&	41.119(42.661)\\ \hline
			-0.1&	196.384(193.942)&	192.319(192.839)&	167.951(169.466)&	141.627(139.787)&	140.180(142.170)\\ \hline
			-0.08&	227.430(230.954)&	221.993(212.506)&	204.823(202.863)&	184.101(185.665)&	176.617(171.668)\\ \hline
			-0.06	&238.217(232.866)&	255.079(251.453)	&262.418(260.889)&	225.436(226.483)&	213.765(214.318)\\ \hline
			-0.04	&245.579(249.019)&	281.759(289.037)&	296.356(297.206)&	272.473(280.987)	&299.414(295.552)\\ \hline
			0	&264.461(260.016)	&302.359(299.338)&	347.475(346.286)&	325.207(320.825)&	305.974(305.671)\\ \hline
			0.04&	260.555(266.668)&	270.862(269.939)&	302.0707(299.2512)&	310.852(313.693)&	318.290(314.839)\\ \hline
			0.06&	251.870(262.379) &	250.298(254.724)&	270.005(272.069)&	275.999(281.701)&	272.290(270.187)\\ \hline
			0.08	&244.245(245.182) 	&219.787(219.027)&	234.629(237.815)&	238.589(238.558)&	250.700(251.219)\\ \hline
			0.1	 &232.486(231.131)&	199.192(196.576)	&199.626(199.655)&	208.923(205.774)&	225.847(231.962)\\ \hline
			0.2	&167.308(169.207)&	117.223(122.804)&	90.241(93.363)&	90.194(89.626)&	102.355(103.080)\\ \hline
			0.3	&120.065(119.569) &	69.55(71.342) &	45.143(46.007)	&42.845(42.373)&	52.055(53.828)\\ \hline

			\multicolumn{6}{|c|}{\(\mathrm{\Delta}\theta_1 = 0\)} \\
			\hline
			-0.3	&33.843(34.723)	&18.083(18.055)	&9.359(9.918)	&9.041(9.484)&	11.411(12.092)\\ \hline
			-0.2&	82.916(85.163)&	55.870(56.069)	&34.607(33.762)	&30.926(31.430)&	36.350(36.493)\\ \hline
			-0.1	&188.537(197.145)	& 171.350(169.514)	&132.168(132.017) &	122.205(120.643)	& 124.282(119.596)\\ \hline
			-0.08	&217.056(215.485)&	203.912(198.949)&	178.908(175.220)&	159.795(161.929)	&162.844(164.547)\\ \hline
			-0.06	&241.340(239.375)&	237.240(236.823)&	236.547(236.814)&	199.675(202.473)&	193.226(195.279)\\ \hline
			-0.04	&259.170(259.076)	&277.023(274.939)	&284.949(289.631)&	253.918(251.700)&	234.745(231.929)\\ \hline
			0.04&	312.479(317.915)&	323.701(323.925)&	350.867(352.415)&	344.498(342.401)&	337.710(36.258)\\ \hline
			0.06&	293.990(297.279)&	314.947(314.624)	&333.444(334.640)&	337.948(332.171)&	330.082(333.784)\\ \hline
			0.08	&274.404(278.698)	&291.258(291.695)	&316.944(309.851)&	306.533(310.780)	&305.513(307.452)\\ \hline
			0.1	&250.737(251.068)&	266.244(274.098)&	266.315(269.102)&	264.719(254.531)&	270.681(271.645)\\ \hline
			0.2	&212.966(213.707)&	155.478(156.436)&	122.496(121.412)&	119.200(118.801) &	127.151(128.334)\\ \hline
			0.3	&158.448(158.556)	&90.728(91.651)	&56.040(56.104)	&54.003(56.396)&	61.223(62.157)\\ \hline

			\multicolumn{6}{|c|}{\(\mathrm{\Delta}\theta_1 = 0.02\)} \\
			\hline
			-0.3&	27.712(29.085)&	15.048(16.021)	&8.183(8.795)&	8.098(8.515)&	10.631(10.168)\\ \hline
			-0.2	&70.240(70.721)	&44.399(45.039)&	27.954(28.436)	&28.406(28.793)&	32.891(33.428)\\ \hline
			-0.1	&167.806(169.762)	&138.154(134.827)	&109.975(108.54)&	105.876(105.607)&	112.874(112.813)\\ \hline
			-0.08&	196.734(196.508)&	169.530(167.099)&	144.911(144.463)&	134.039(134.799)&	143.418(138.816)\\ \hline
			
			-0.06&	223.883(221.746)	&213.914(213.077)&	195.114(197.595)&	179.273(178.847)	&178.390(178.101)\\ \hline
			-0.04&	239.528(232.141)&	257.109(258.640)&	245.931(247.635)&	222.773(223.727)	&225.379(227.734)\\ \hline
			0	& 308.056(307.786)&	339.750(341.465)&	354.925(953.472)&	339.670(345.077)	&325.626(322.046)\\ \hline
			0.04	&319.748(318.980)	&324.168(323.698)&	346.189(344.025)&	327.068(328.146)	&321.598(322.068)\\ \hline
			0.06&	312.473(310.297)&	315.684(314.006)&	332.068(334.147)&	311.427(313.658)&	300.268(302.145)\\ \hline
			0.08	& 289.663(291.542)	& 298.164(300.100)	& 324.158(326.860)&	301.698(303.106)	&289.998(290.476)\\ \hline
			0.1	&245.371(248.058)	&264.151(266.138)	&281.120(282.67)&	267.169(263.008)&	244.689(245.138)\\ \hline
			0.2	&224.807(228.808)&	239.148(240.168)	&251.069(253.132)&	236.292(237.648)&	218.118(219.036)\\ \hline
			0.3	&200.292(199.285)&	212.069(214.008)&	225.149(227.348)&	206.448(207.369)&	195.107(199.147)
			 \\ \hline
			\multicolumn{6}{|c|}{\(\mathrm{\Delta}\theta_2 = - 0.02\)} \\
			\hline
			$\Delta\theta_1$	& $ARL_1(SDRL_1)$ & $ARL_1(SDRL_1)$ & $ARL_1(SDRL_1)$ & $ARL_1(SDRL_1)$ & $ARL_1(SDRL_1)$\\ \hline
			-0.3	& 10.840(11.114) &	6.885(7.413) &	9.441(9.895) &	17.123(17.186)&	31.322(31.134)\\ \hline
			-0.2	& 35.597(36.154)&	24.420(25.439) &	36.444(37.566)&	57.874(58.542)	& 85.959(85.056)\\ \hline
			-0.1 &	126.590(126.240) &	102.723(103.322) &	155.050(155.213) &	190.384(188.427) &	213.890(211.796) \\ \hline
			-0.08 &	156.695(157.751) &	133.621(133.328) &	205.551(204.911) &	229.128(230.449) &	242.053(242.826)\\ \hline
			-0.06  &	194.156(199.628)	 & 178.783(177.285)	 & 263.096(269.806)&	272.024(269.430) &	265.992(266.505)\\ \hline
			-0.04 &	231.584(230.611) &	227.591(230.098) &	306.494(310.377) &	296.610(291.043) &	276.81(284.007) \\ \hline
			0	& 276.364(282.367) &	324.399(317.740) &	327.217(335.156) &	303.375(305.815)&	281.545(2276.686)\\ \hline
			0.04 &	258.763(264.191) &	344.873(345.091) &	260.653(259.653)&	254.561(252.305) &	253.287(263.549)\\ \hline
			0.06	 & 238.850(236.982)	 & 322.309(325.219)	 & 223.197(226.037)&	230.649(231.182) &	233.761(233.284)\\ \hline
			0.08	 & 211.070(211.663) &	290.066(292.107) &	184.931(186.109)&	200.765(204.647) &	214.624(217.458)\\ \hline
			0.1	& 174.956(174.676) &	233.503(237.820) &	148.670(149.913)&	164.991(169.717) &	189.084(194.387)\\ \hline
			0.2	 &68.472(68.372)&	85.631(85.583) &	53.014(52.938) &	77.701(77.689)&	109.839(112.508)\\ \hline
			0.3	& 31.126(31.305) & 34.895(35.694) &	21.788(22.365) &	38.119(38.438)	&61.199(61.199)\\ \hline
			
			\multicolumn{6}{|c|}{\(\mathrm{\Delta}\theta_2 = 0\)} \\
			\hline
			-0.3 &	10.034(10.381) &	7.647(8.272)&	8.302(8.669) &	15.857(16.567) &24.369(25.017)	\\ \hline
			-0.2 &	33.251(35.058) &	27.951(29.210) &	31.427(32.405)&	69.058(71.401) &	74.826(76.246)\\ \hline
			-0.1 &	118.095(118.319) &	111.605(109.883) &	 133.201(136851) &	171.067(172.399)&	195.372(193.879)\\ \hline
			-0.08 &	146.249(146.090) &	153.231(158.601) &	183.685(181.624) &	205.787(209.753) &	228.709(230.133)\\ \hline
			-0.06 &	185.654(183.334) &	200.141(203.143) &	234.462(237.095) &	248.910(247.287) &	261.094(267.136)\\ \hline
			-0.04 &	223.385(223.601) &	249.392(250.315) &	285.440(280.372) &	296.775(297.363) &	289.310(289.173) \\ \hline
			0.04  &	291.136(287.433) &	314.862(319.323) &	332.790(327.014) &	304.035(313.737) &	308.516(307.032)\\ \hline
			0.06 &	272.882(276.039) &	280.267(276.925) &	 290.301(295.216) &	288.940(289.684) &	287.980(287.086)\\ \hline
			0.08 &	242.103(240.841) &	241.573(240.860) &	242.365(245.558) &	245.994(246.845) &	270.146(266.655)\\ \hline 
			0.1	&205.654(205.491)&	191.214(187.732)&	195.658(194.166) &	216.483(216.234) &	239.941(246.096)\\ \hline
			0.2	&86.649(89.245)	&71.207(71.905)	 &78.253(71.258) &102.873(103.785)&	140.481(140.108)\\ \hline
			0.3	&36.442(37.068)	&29.168(27.169)	&27.545(27.821)	&47.448(47.748)&	79.358(80.306)\\ \hline
			
			\multicolumn{6}{|c|}{\(\mathrm{\Delta}\theta_2 = 0.02\)} \\
			\hline
			-0.3	& 9.650(10.284)	 &7.299(7.958)	 &0.589(1.033)	 &13.477(14152)	&23.687(24.461)\\ \hline
			-0.2	&31.202(31.115)	 &24.683(24.788)	 &28.376(28.861)&	43.082(44.009)	 &64.699(64.033)\\ \hline
			-0.1	&108.593(109.140) &	101.083(105.358) &	118.527(117.349)&	146.757(146.798) &	174.042(174.954)\\ \hline
			-0.08	 &137.486(136.332) &	135.200(135.297) &	148.392(150.669) &	182.224(184.964) &	211.310(212.253)\\ \hline
			-0.06	&179.196(186.333)	& 176.150(175.810)	& 207.790(210.484)&	233.010(237.697) &	248.803(243.346)\\ \hline
			-0.04	 &238.871(237.194)	& 230.519(233.929)	 &265.570(260.090) &	270.982(273.964) &	278.320(274.913)\\ \hline
			0	& 308.145(310.576)	 & 326.215(326.027)	 & 341.309(342.787)&	335.724(336.217)	&337.572(340.536)\\ \hline
			0.04&	307.018(309.323)	&325.257(323.759)	& 359.427(361.850)&	343.987(341.803)&	330.710(337.482)\\ \hline
			0.06	&288.145(285.073)	&303.641(309.915)	&327.819(338.975)& 	323.327(324.212)	&325.134(326.386)\\ \hline
			0.08	& 257.106(252.615)	&266.973(267.113)	&312.147(313.708)&	300.053(302.597) &	312.819(313.720)\\ \hline
			0.1	& 218.869(220.067)	&225.369(224.655)& 	244.279(241.177)&	270.159(277.502)&	281.247(284.837)\\ \hline
			0.2	 & 88.373(91.724)	& 98.113(98.602) &	91.526(90.171)&	102.658(104.298) &	116.845(117.547)\\ \hline
			0.3	 &33.051(32.628) &	46.233(45.738)	& 37.097(37.29)	& 42.454(41.989)&	49.325(52.497)\\ \hline
			
	\end{longtable}}
\end{center}

\begin{center}
	{\tiny 
	\begin{longtable}{|c|c|c|c|c|c|}
		\caption{OOC performance of Tbeta Percentiles for $\theta_1=2, \theta_2=15,  $a=0.1$,  $b=0.6$,  \nu=0.0027$}\label{tab6}\\
		\hline
		& $p=0.1$ & $p=0.25$ & $p=0.5$ & $0.75$ & $0.9$\\
		\hline
		\endfirsthead
		\multicolumn{6}{l}%
		{\tablename\ \thetable\ -- \textit{Continued from previous page}} \\
		\hline
		$\Delta\theta_2$ & $p=0.1$ & $p=0.25$ & $p=0.5$ & $0.75$ & $0.9$\\
		\hline
		\endhead
		\hline \multicolumn{6}{r}{\textit{Continued on next page}} \\
		\endfoot
		\hline
		\endlastfoot
		\multicolumn{6}{|c|}{\(\mathrm{\Delta}\theta_1 = -0.02\)} \\
		\hline
		$\Delta\theta_2$	& $ARL_1(SDRL_1)$ & $ARL_1(SDRL_1)$ & $ARL_1(SDRL_1)$ & $ARL_1(SDRL_1)$ & $ARL_1(SDRL_1)$\\
		\hline
				-0.3 & 47.859 (47.310) & 25.378 (25.412) & 18.369(17.036) & 14.036(15.028) & 11.0358(10.014) \\ \hline
			-0.2 & 116.031 (116.424) & 74.416 (73.340) & 48.079 (47.558) & 38.372 (39.495) & 35.683 (35.293) \\ \hline
			-0.1 & 259.176 (262.491) & 208.854 (206.853) & 182.526 (182.802) & 152.426 (145.862) & 119.191 (119.753) \\ \hline
			-0.08 & 297.346 (299.477) & 261.093 (275.553) & 226.717 (226.987) & 197.733 (201.652) & 148.211 (148.755) \\ \hline
			-0.06 & 322.346 (322.639) & 300.567 (311.454) & 276.723 (277.521) & 242.959 (246.572) & 190.166 (189.776) \\ \hline
			-0.04 & 344.106 (346.919) & 328.096 (329.664) & 328.002 (328.713) & 297.117 (296.783) & 233.020 (238.484) \\ \hline
			0 & 355.785 (358.819) & 375.316 (365.522) & 379.292 (397.958) & 368.036(370.002) & 323.068(325.068) \\ \hline
			0.04 & 344.900 (345.937) & 345.092 (347.900) & 347.339 (355.984) & 362.437 (358.009) & 307.316 (308.935) \\ \hline
			0.06 & 311.625 (313.800) & 328.644 (331.676) & 325.207 (324.114) & 337.542 (337.969) & 310.712 (308.935) \\ \hline
			0.08 & 279.518 (279.517) & 283.399 (285.668) & 292.324 (283.685) & 295.576 (303.797) & 276.991 (270.296) \\ \hline
			0.1 & 250.949 (254.352) & 257.449 (263.230) & 252.353 (253.901) & 260.230 (273.508) & 247.704 (252.687) \\ \hline
			0.2 & 134.268 (135.537) & 131.998 (133.145) & 130.195 (134.046) & 132.943 (134.495) & 129.661 (133.917) \\ \hline
			0.3 & 70.607 (70.202) & 72.329 (74.219) & 62.0582(34.682) & 69.328(71.069) & 59.038(61.039)\\ \hline
			\multicolumn{6}{|c|}{\(\mathrm{\Delta}\theta_1 = 0\)} \\
			\hline
			-0.3 & 43.151 (43.547) & 23.942 (24.762) & 19.036(18.679) & 26.168(25.068) & 19.358(19.038)\\ \hline
			-0.2 & 101.494 (101.140) & 68.302 (68.237) & 42.361 (42.246) & 35.603 (36.525) & 32.933 (33.568) \\ \hline
			-0.1 & 239.702 (237.867) & 191.625 (191.796) & 155.284 (154.078) & 138.882 (140.882) & 110.000 (111.742) \\ \hline
			-0.08 & 260.987 (257.918) & 233.721 (229.322) & 254.048 (260.211) & 172.628 (175.286) & 138.289 (135.607) \\ \hline
			-0.06 & 311.889 (312.667) & 280.664 (278.422) & 245.618 (246.783) & 217.370 (219.152) & 171.263 (175.878) \\ \hline
			-0.04 & 330.856 (336.178) & 319.057 (320.764) & 309.611 (311.331) & 266.662 (268.914) & 212.238 (213.110) \\ \hline
			0.04 & 352.734 (354.917) & 367.473 (364.538) & 394.850 (391.537) & 373.248 (373.979) & 312.417 (316.223) \\ \hline
			0.06 & 320.548 (295.680) & 340.542 (339.337) & 360.657 (364.123) & 352.105 (358.078) & 317.721 (316.479) \\ \hline
			0.08 & 293.568 (295.686) & 319.734 (320.629) & 322.851 (320.055) & 330.903 (339.457) & 304.427 (302.607) \\ \hline
			0.1 & 276.849 (281.168) & 280.634 (276.928) & 288.147 (274.208) & 279.304 (278.573) & 270.080 (269.348) \\ \hline
			0.2 & 147.498 (147.292) & 150.511 (152.842) & 143.517 (141.089) & 144.538 (148.920) & 141.708 (141.457) \\ \hline
			0.3 & 78.687 (81.236) & 78.014 (81.041) & 72.321(69.518) & 77.130(78.367) & 82.037(80.157)\\ \hline
			
			\multicolumn{6}{|c|}{\(\mathrm{\Delta}\theta_1 = 0.02\)} \\
			\hline
			-0.3 & 37.322 (36.761) & 20.951 (20.216) & 12.0364(11.358) & 13.878(2.119)& 18.369(19.038)\\ \hline
			-0.2 & 94.928 (96.742) & 61.423 (60.346) & 38.035 (37.396) & 31.007 (32.050) & 30.060 (30.371) \\ \hline
			-0.1 & 212.928 (213.363) & 176.353 (179.010) & 138.991 (139.382) & 122.573 (121.075) & 101.429 (99.178) \\ \hline
			-0.08 & 249.737 (244.094) & 209.044 (205.560) & 175.617 (179.842) & 157.375 (159.616) & 132.193 (130.966) \\ \hline
			-0.06 & 281.960 (277.163) & 249.840 (249.927) & 226.466 (232.365) & 203.347 (207.421) & 156.372 (157.273) \\ \hline
			-0.04 & 314.814 (314.234) & 300.595 (298.992) & 276.966 (276.634) & 256.328 (257.358) & 194.428 (186.955) \\ \hline
			0 & 359.150 (352.720) & 371.266 (370.733) & 378.262(379.530) & 372.024(371.880) & 359.067(358.067) \\ \hline
			0.04 & 335.152 (333.213) & 359.549 (359.294) & 376.162 (373.927) & 375.101 (379.592) & 330.593 (330.687) \\ \hline
			0.06 & 325.127 (321.634) & 349.828 (348.424) & 349.738 (350.497) & 365.632 (368.788) & 327.276 (324.672) \\ \hline
			0.08 & 318.267 (313.421) & 328.541 (327.352) & 339.987 (347.100) & 353.884 (352.200) & 322.115 (312.561) \\ \hline
			0.1 & 302.343 (301.191) & 297.953 (303.151) & 315.619 (315.043) & 321.677 (321.997) & 286.071 (284.076) \\ \hline
			0.2 & 159.182 (160.328) & 151.263 (152.069) & 153.566 (154.453) & 166.462 (168.965) & 164.373 (172.627) \\ \hline
			0.3 & 86.286 (86.578) & 84.460 (86.453) & 75.659(78.036) & 68.269(69.258) & 64.259(65.238) \\ \hline
			\multicolumn{6}{|c|}{\(\mathrm{\Delta}\theta_2 = -0.02\)} \\
			\hline
			$\Delta\theta_1$	& $ARL_1(SDRL_1)$ & $ARL_1(SDRL_1)$ & $ARL_1(SDRL_1)$ & $ARL_1(SDRL_1)$ & $ARL_1(SDRL_1)$\\ \hline
			-0.3 & 151.434 (153.592) & 156.565 (158.617) & 162.325(163.359) & 169.035(170.247) & 145.268(144.268) \\ \hline
			-0.2 & 236.006 (236.951) & 245.999 (246.678) & 255.801 (254.250) & 239.362 (236.165) & 213.595 (216.713) \\ \hline
			-0.1 & 330.331 (333.184) & 340.175 (338.183) & 346.049 (345.895) & 335.381 (343.537) & 260.682 (260.116) \\ \hline
			-0.08 & 347.010 (342.662) & 365.604 (362.121) & 362.679 (359.408) & 340.973 (338.209) & 270.727 (273.437) \\ \hline
			-0.06 & 369.503 (367.318) & 359.897 (361.178) & 343.016 (342.804) & 337.173 (330.988) & 273.084 (266.537) \\ \hline
			-0.04 & 375.137 (386.206) & 366.910 (366.182) & 376.510 (368.262) & 349.035 (337.387) & 269.772 (272.932) \\ \hline
			0 & 358.653 (368.402) & 346.344 (348.197) & 334.269(335.068) & 336.297(338.169) & 329.036(328.076) \\ \hline
			0.04 & 333.077 (340.003) & 321.178 (320.521) & 293.708 (294.915) & 288.315 (285.301) & 231.257 (227.916) \\ \hline
			0.06 & 308.944 (312.639) & 295.375 (288.033) & 294.267 (302.412) & 260.155 (267.806) & 213.749 (215.492) \\ \hline
			0.08 & 290.575 (290.110) & 267.101 (268.825) & 262.932 (284.825) & 242.361(240.165) & 205.353 (199.546) \\ \hline
			0.1 & 266.716 (266.685) & 249.370 (248.433) & 237.585 (241.392) & 221.036(218.469) & 189.619 (185.890) \\ \hline
			0.2 & 167.042 (165.737) & 141.705 (143.359) & 137.957 (142.530) & 126.349(124.268) & 125.556 (125.736) \\ \hline
			0.3 & 105.474 (105.714) & 82.752 (81.929) & 74.368(77.126) & 68.127(69.249) & 59.260(61.230) \\
			\hline
				\multicolumn{6}{|c|}{\(\mathrm{\Delta}\theta_2 = 0\)} \\
			\hline
			-0.3 & 133.746 (136.748) & 142.511 (144.269) & 145.169(144.358) & 138.669(141.358) & 105.333(106.449)\\ \hline
			-0.2 & 210.177 (215.861) & 220.910 (220.972) & 222.228 (227.163) & 218.730 (220.001) & 202.912 (201.012) \\ \hline
			-0.1 & 310.286 (307.428) & 323.910 (323.485) & 337.288 (330.394) & 324.791 (328.494) & 283.245 (278.012) \\ \hline
			-0.08 & 324.480 (326.720) & 344.727 (347.512) & 344.723 (341.659) & 330.568 (330.262) & 282.479 (276.903) \\ \hline
			-0.06 & 340.592 (343.056) & 356.368 (358.345) & 359.979 (359.382) & 346.897 (345.507) & 290.018 (287.167) \\ \hline
			-0.04 & 359.774 (354.366) & 355.918 (356.219) & 376.669 (380.892) & 356.183 (357.929) & 293.719 (290.211) \\ \hline
			0.04 & 350.265 (347.661) & 347.157 (352.566) & 349.815 (352.433) & 349.147 (348.980) & 271.487 (271.396) \\ \hline
			0.06 & 333.072 (332.581) & 330.404 (332.302) & 337.186 (340.393) & 316.551 (312.691) & 251.188 (246.601) \\ \hline
			0.08 & 326.165 (330.691) & 312.128 (320.521) & 311.853 (315.709) & 305.376 (293.958) & 244.578 (214.236) \\ \hline
			0.1 & 298.276 (322.275) & 249.370 (248.433) & 291.175 (293.649) & 282.641 (290.065) & 227.222 (222.657) \\ \hline
			0.2 & 200.805 (199.388) & 141.705 (143.359) & 170.235 (169.408) & 175.560 (177.329) & 142.888(145.928) \\ \hline
			0.3 & 125.490 (125.126) & 104.277 (103.193) & 131.267(132.069) & 144.267(141.359) & 152.787(151.795) \\ \hline
			\multicolumn{6}{|c|}{\(\mathrm{\Delta}\theta_2 = 0.02\)} \\
			\hline
			-0.3 & 117.640 (115.863) & 118.398 (119.233) & 123.558(124.568) & 128.698(129.349) & 122.568(121.269) \\ \hline
			-0.2 & 184.649 (178.912) & 194.545 (190.351) & 193.059 (188.802) & 192.887 (198.410) & 184.755 (183.414) \\ \hline
			-0.1 & 292.687 (293.865) & 297.821 (291.856) & 304.019 (305.279) & 297.092 (299.044) & 260.839 (271.025) \\ \hline
			-0.08 & 311.978 (310.295) & 324.259 (317.556) & 329.895 (328.082) & 325.480 (324.433) & 295.307 (301.210) \\ \hline
			-0.06 & 328.256 (326.268) & 331.348 (342.132) & 344.315 (344.459) & 347.024 (353.170) & 296.184 (289.332) \\ \hline
			-0.04 & 337.229 (333.731) & 347.895 (342.239) & 362.615 (368.296) & 356.388 (348.456) & 292.608 (292.889) \\ \hline
			0 & 369.835 (365.568) & 371.534 (367.196) & 372.692(373.277) & 375.269(374.469) & 356.358(358.369) \\ \hline
			0.04 & 358.824 (355.801) & 359.611 (367.136) & 368.043 (365.615) & 366.294 (368.321) & 290.339 (287.282) \\ \hline
			0.06 & 344.730 (351.900) & 360.484 (355.097) & 365.107 (368.607) & 360.094 (359.298) & 287.800 (291.136) \\ \hline
			0.08 & 340.557 (333.759) & 357.438 (365.965) & 347.435 (356.357) & 338.268(340.446) & 288.270 (281.302) \\ \hline
			0.1 & 334.188 (328.026) & 351.926 (357.438) & 341.670 (341.784) & 336.497(338.169) & 266.955 (274.482) \\ \hline
			0.2 & 233.759 (246.301) & 218.261 (216.764) & 217.242 (215.810) & 223.778(224.697) & 193.954 (196.693) \\ \hline
			0.3 & 140.704 (141.431) & 129.329 (130.315) & 135.696(133.448) & 150.300(152.349) & 112.369(115.465) \\ \hline
			
	\end{longtable}}
\end{center}

\subsection{OOC Performance Comparison}
\hspace*{0.2in} The OOC performance of the $SBTBC$ chart is compared with bootstrap beta control chart as proposed by Chowdhury \emph{et al.} (2022) assuming process to follow beta distribution. For $\nu=0.0027,$ the comparisons are carried out for different choices of shifts ($\Delta\theta_1$ and/or $\Delta\theta_2$) in the MLEs of the shape parameters of both beta and Tbeta distribution. The OOC results for the bootstrap beta control chart are exhibited in Table $\ref{tab7}$. Comparing the OOC results from Tables $\ref{tab5}$ and $\ref{tab7}$, and from Figures \ref{figure15}-\ref{figure17}, it can be claimed that $SBTBC$ chart detects any type of shift ($\Delta\theta_1$ and/or $\Delta\theta_2$) of any magnitude (small, moderate or large) faster than the bootstrap beta control chart.

\begin{center}
{\tiny
	\begin{longtable}{|c|c|c|c|c|c|}
		\caption{OOC performance  of beta percentiles for $\theta_1=2, \theta_2=15, \nu=0.0027$}\label{tab7}\\
		\hline
		 & $p=0.1$ & $p=0.25$ & $p=0.5$ & $0.75$ & $0.9$\\
		\hline
		\endfirsthead
		\multicolumn{6}{l}%
		{\tablename\ \thetable\ -- \textit{Continued from previous page}} \\
		\hline
		 & $p=0.1$ & $p=0.25$ & $p=0.5$ & $0.75$ & $0.9$\\
		\hline
		\endhead
		\hline \multicolumn{6}{r}{\textit{Continued on next page}} \\
		\endfoot
		\hline
		\endlastfoot
		\multicolumn{6}{|c|}{$\Delta\theta_1=-0.02$}\\
		\hline
	 $\Delta\theta_2$	& $ARL_1(SDRL_1)$ & $ARL_1(SDRL_1)$ & $ARL_1(SDRL_1)$ & $ARL_1(SDRL_1)$ & $ARL_1(SDRL_1)$\\ \hline
		-0.3	& 37.722(38.217) 	& 19.895(20.479) 	& 11.854(12.557) 	& 9.414(9.798) 	& 13.490(14.209) \\ \hline
		-0.2	& 91.808(90.603) 	& 60.687(60.154) 	& 40.749(41.774) 	& 35.824(35.248) 	& 49.337(49.292) \\ \hline
		-0.1	& 207.757(210.764) 	& 183.605(186.394) 	& 154.284(152.866) 	& 147.014(148.974) 	& 180.319(184.304) \\ \hline
		-0.08	& 235.164(237.831) 	& 215.774(212.754) 	& 196.206(196.735) 	& 188.399(188.813) 	& 213.802(211.426) \\ \hline
		-0.06	& 264.823(261.887) 	& 263.182(264.622) 	& 242.553(246.160) 	& 241.754(237.754) 	& 266.348(270.612) \\ \hline
		-0.04	& 300.229(310.583) 	& 291.825(289.604) 	& 296.919(300.640) 	& 299.085(280.972) 	& 300.348(293.037) \\ \hline
		0	& 347.471(352.078) 	& 335.169(340.003) 	& 358.285(357.582) 	& 349.160(352.160) 	& 336.429(329.211) \\ \hline
		0.04	& 314.538(319.250) 	& 303.348(306.409) 	& 325.219(328.830) 	& 339.967(342.966) 	& 325.256(319.939) \\ \hline
		0.06	& 303.028(307.927) 	& 287.019(286.998) 	& 307.609(309.057) 	& 325.446(320.957) 	& 295.372(289.502) \\ \hline
		0.08	& 275.040(272.258) 	& 258.861(259.419) 	& 269.450(268.345) 	& 297.483(300.785) 	& 264.892(268.559) \\ \hline
		0.1	& 238.136(237.838) 	& 222.731(223.264) 	& 240.426(236.006) 	& 265.474(273.737) 	& 231.996(231.555) \\ \hline
		0.2	& 185.878(487.351) 	& 135.991(137.295) 	& 109.983(106.989) 	& 125.210(126.188) 	& 109.889(113.834) \\ \hline
		0.3	& 126.381(122.643) 	& 85.503(84.633) 	& 54.379(55.120) 	& 62.843(63.843) 	& 53.876(53.964) \\ \hline
		
		\multicolumn{6}{|c|}{$\Delta\theta_1=0$}\\
		\hline
		-0.3	& 32.117(32.327) 	& 16.417(17.083) 	& 9.406(10.124) 	& 8.167(8.794) 	& 12.822(13.554) \\ \hline
		-0.2	& 79.342(79.898) 	& 51.512(52.115) 	& 34.258(34.276) 	& 30.787(31.911) 	& 44.209(45.442) \\ \hline
		-0.1	& 183.901(183.210) 	& 174.210(176.374) 	& 122.643(122.433) 	& 126.212(128.602) 	& 155.216(158.878) \\ \hline
		-0.08	& 209.677(210.332) 	& 217.781(216.988) 	& 166.236(162.762) 	& 168.863(168.179) 	& 204.555(197.150) \\ \hline
		-0.06	& 240.915(241.188) 	& 247.527(247.136) 	& 215.691(226.901) 	& 212.751(213.280) 	& 244.382(243.969) \\ \hline
		-0.04	& 267.571(270.649) 	& 280.325(282.187) 	& 256.861(255.564) 	& 271.345(267.080) 	& 293.067(292.770) \\ \hline
		0.04	& 325.646(330.623) 	& 338.292(329.240) 	& 369.968(373.127) 	& 355.430(358.427) 	& 369.633(367.299) \\ \hline
		0.06	& 306.312(307.433) 	& 327.413(326.171) 	& 360.166(362.955) 	& 341.016(342.314) 	& 355.266(356.202) \\ \hline
		0.08	& 281.909(283.044) 	& 306.380(308.713) 	& 339.498(342.809) 	& 327.624(325.090) 	& 342.460(343.955) \\ \hline
		0.1	& 246.478(247.228) 	& 288.243(289.662) 	& 290.101(290.101) 	& 281.170(289.750) 	& 328.541(329.402) \\ \hline
		0.2	& 227.757(232.464) 	& 192.830(191.674) 	& 144.757(142.808) 	& 158.775(158.414) 	& 195.657(198.186) \\ \hline
		0.3	& 192.211(193.404) 	& 105.292(106.797) 	& 70.422(75.011) 	& 78.192(79.902) 	& 95.808(96.223) \\ \hline
		
		\multicolumn{6}{|c|}{$\Delta\theta_1=0.02$}\\
		\hline
		-0.3	& 26.848(27.161) 	& 44.041(44.527) 	& 18.099(18.585) 	& 6.761(7.252) 	& 11.310(11.724) \\ \hline
		-0.2	& 68.655(70.545) 	& 82.030(82.125) 	& 38.176(38.499) 	& 25.119(26.119) 	& 38.519(39.239) \\ \hline
		-0.1	& 157.599(158.848) 	& 168.711(170.553) 	& 115.079(116.078) 	& 109.769(107.841) 	& 137.755(137.107) \\ \hline
		-0.08	& 184.732(185.602) 	& 189.454(184.178) 	& 157.090(152.996) 	& 142.103(140.917) 	& 181.673(183.901) \\ \hline
		-0.06	& 210.038(219.851) 	& 235.414(236.734) 	& 208.274(202.641) 	& 188.115(186.289) 	& 230.005(232.701) \\ \hline
		-0.04	& 244.316(247.806) 	& 272.378(270.416) 	& 278.987(278.221) 	& 241.550(242.857) 	& 293.834(292.445) \\ \hline
		0	& 308.770(305.791) 	& 354.088(356.777) 	& 368.769(369.417) 	& 351.384(354.330) 	& 338.003(338.513) \\ \hline
		0.04	& 334.465(337.512) 	& 347.724(347.897) 	& 362.977(369.455) 	& 357.726(352.434) 	& 331.382(332.583) \\ \hline
		0.06	& 319.850(326.199) 	& 327.047(327.953) 	& 343.059(341.138) 	& 352.072(350.947) 	& 317.259(315.199) \\ \hline
		0.08	& 299.191(294.581) 	& 307.432(302.162) 	& 327.611(331.773) 	& 324.443(325.975) 	& 361.048(368.248) \\ \hline
		0.1	& 261.383(263.819) 	& 282.091(282.203) 	& 305.574(303.945) 	& 296.499(295.168) 	& 284.578(287.029) \\ \hline
		0.2	& 241.579(242.823) 	& 256.839(255.138) 	& 278.327(275.107) 	& 241.286(246.546) 	& 232.412(239.886) \\ \hline
		0.3	& 218.764(214.542) 	& 232.547(228.142) 	& 243.931(241.228) 	& 216.388(217.643) 	& 219.476(218.893) \\ \hline
		\multicolumn{6}{|c|}{$\Delta\theta_2=-0.002$}\\
		\hline
		 $\Delta\theta_1$	& $ARL_1(SDRL_1)$ & $ARL_1(SDRL_1)$ & $ARL_1(SDRL_1)$ & $ARL_1(SDRL_1)$ & $ARL_1(SDRL_1)$\\ \hline
		-0.3 & 38.293(49.106) & 14.914(14.346) & 34.182(34.507) & 21.320(22.033)& 32.131(31.966)\\ \hline
		-0.2 & 43.249(41.192) & 43.137(45.460) & 56.873(51.589) & 74.652(74.821)&  86.507(87.730)\\ \hline
		-0.1 & 172.963(174.088) & 121.778(124.981) & 180.414(183.603) & 230.443(230.821)& 221.966(218.497)\\ \hline
		-0.08 & 214.774(214.457) & 168.319(165.822) & 219.118(218.821) & 277.458(273.123)&261.351(263.405) \\ \hline
		-0.06 & 352.451(345.299) & 194.370(190.222) & 291.012(295.768) & 297.671(300.183)& 293.955(265.579)\\ \hline
		-0.04 & 282.820(286.323) & 261.694(263.388) & 316.731(313.615) & 323.441(313.009)& 307.230(308.011)\\ \hline
		0 & 300.846(297.542) & 362.269(365.351) & 315.261(315.516) & 332.875(330.573) &344.473(346.856) \\ \hline
		0.04 & 260.185(262.056) & 353.804(353.04) & 281.067(283.546) & 267.754(271.061)& 326.547(330.643)\\ \hline
		0.06 & 252.253(253.371) & 343.028(342.011) & 242.052(243.928) & 241.387(236.711)& 307.785(303.753)\\ \hline
		0.08 & 234.175(231.508) & 313.966(313.699) & 208.085(204.042) & 223.646(225.899)& 290.722(290.766)\\ \hline
		0.1 & 193.384(193.105) & 241.286(242.618) & 179.316(180.391) & 192.092(190.050)& 250.556(254.181)\\ \hline
		0.2 & 69.810(70.081) & 95.807(96.515) & 53.474(53.288) & 73.280(72.284)& 154.896(155.034)\\ \hline
		0.3 & 31.255(31.939) & 22.306(23.035) & 23.416(24.005) & 36.202(36.692)& 88.628(88.346)\\ \hline

		\multicolumn{6}{|c|}{$\Delta\theta_2=0$}\\
		\hline
		-0.3 & 38.343(36.535) & 36.059(35.903) & 31.744(13.436) & 18.586(19.384)& 27.436(27.588)\\ \hline
		-0.2 & 35.386(32.945) & 41.734(41.518) & 52.241(50.228) & 163.092(163.280)& 75.913(75.014)\\ \hline
		-0.1 & 143.723(147.908) & 193.849(195.911) & 203.303(204.378) & 233.843(238.253)&197.657(195.484)\\ \hline
		-0.08 & 238.507(239.285) & 294.462(297.247) & 245.631(249.300) & 253.624(257.193)&239.163(238.525)\\ \hline
		-0.06 & 258.550(259.812) & 249.371(251.245) & 289.906(288.141) & 294.139(298.866)&277.219(278.147)\\ \hline
		-0.04 & 268.445(263.639) & 283.203(284.1777) & 313.831(314.824) & 340.635(341.651)&322.485(320.892)\\ \hline
		0.04 & 287.561(287.250) & 338.737(340.530) & 348.621(349.275) & 342.803(346.241)&371.581(370.769)\\ \hline
		0.06 & 278.196(278.919) & 311.351(313.516) & 318.237(319.514) & 308.354(307.300)&332.687(333.642)\\ \hline
		0.08 & 265.119(268.404) & 247.723(249.993) & 250.143(253.659) & 263.196(272.259)&292.633(290.054)\\ \hline
		0.1 & 228.148(229.913) & 194.559(194.435) & 201.415(205.606) & 230.732(227.666)&276.914(273.951)\\ \hline
		0.2 & 83.495(82.081) & 83.072(85.316) & 91.278(91.761) & 99.469(101.035)& 155.228(158.337)\\ \hline
		0.3 & 37.247(37.881) & 28.503(29.124) & 29.052(29.415) & 45.427(46.874)& 84.913(82.126)\\ \hline
		
		\multicolumn{6}{|c|}{$\Delta\theta_2=0.02$}\\
		\hline
		-0.3 & 39.644(36.565) & 32.153(30.153) & 24.745(26.349) & 16.110(16.242)& 24.884(24.976)\\  \hline
		-0.2 & 59.606(59.411) & 41.730(39.243) & 42.345(49.492) & 55.252(56.600)& 66.850(68.822)\\ \hline
		-0.1 & 144.152(146.805) & 125.693(123.933) & 148.771(150.629) & 183.668(183.396)& 174.771(177.758)\\ \hline
		-0.08 & 169.239(165.272) & 156.148(158.193) & 183.529(186.131) & 228.800(229.020)& 211.844(212.838)\\ \hline
		-0.06 & 235.867(240.315) & 234.615(238.046) & 251.999(250.216) & 280.122(279.440)& 248.554(251.292)\\ \hline
		-0.04 & 302.623(307.210) & 315.590(315.373) & 311.465(311.961) & 328.083(330.048)& 299.461(301.835)\\ \hline
		0 & 324.766(329.147) & 347.787(347.784) & 348.488(349.756) & 356.323(358.174)& 350.501(348.543)\\ \hline
		0.04 & 313.940(309.064) & 334.891(337.750) & 362.173(369.133) & 351.174(352.824)& 329.368(328.170)\\ \hline
		0.06 & 283.587(277.501) & 336.495(334.854) & 328.344(325.054) & 336.373(340.735)&343.406(345.979) \\ \hline
		0.08 & 249.479(254.302) & 295.796(292.582) & 303.958(303.877) & 317.931(314.386)& 328.176(332.385)\\ \hline
		0.1 & 221.643(218.282) & 240.633(242.633) & 266.270(267.386) & 288.305(288.158)& 311.483 (306.046)\\ \hline
		0.2 & 139.059(139.585) & 184.606(183.011) & 141.051(140.405) & 131.530(129.998)& 202.807(201.958)\\ \hline
		0.3 & 64.146(65.171) & 53.905(55.158) & 36.304(36.948) & 60.502(60.391)& 87.855(90.579)\\ \hline
		\end{longtable}}
\end{center}
	
\setcounter{section}{3}
\section{Applications}
\subsection{Monitoring Relative Humidity}
\setcounter{equation}{0}
\hspace*{0.2in} Control charts are valuable tools for monitoring environmental data, allowing for early detection of changes or anomalies in environmental quality indicators over time. They help in distinguishing between natural variability and significant shifts caused by pollution, climate events, or human interventions. For example, control charts can be used to track daily air pollutant concentrations (like $PM2.5$ or $NO_{2}$), monitor water quality parameters (such as $pH$ levels, dissolved oxygen, or heavy metal concentrations), or observe trends in relative humidity or temperature in ecological studies. By signaling when environmental variables exceed acceptable limits, control charts support timely responses and informed policy decisions in environmental management.\\
\hspace*{0.2in} This section demonstrates the application of the $SBTBC$ chart in environmental monitoring, specifically using RH data expressed as proportions. RH is a crucial variable in various domains such as environmental diagnostics and risk assessment (Camuffo, 1998). It significantly influences climatic conditions, including temperature, rainfall, and thermal sensation, and plays a vital role in public health. RH can impact the spread and prevalence of infectious and allergic diseases, and it is particularly important in the transmission of airborne viruses like influenza (Gao \emph{et al.}, 2014). In the context of COVID-19, several studies have underscored the role of RH. For example, Liu \emph{et al.} (2020) and Yao \emph{et al.} (2020) examined how meteorological variables, including RH, influence viral transmission. Sarkodie and Owusu (2020) found that higher temperature and RH reduce the virus’s viability and spread, while Biktasheva (2020) advocated for indoor humidity control as a preventive measure. Yang \emph{et al.} (2021) further showed that RH and temperature are significant drivers of COVID-19 transmission, with effects varying by season and region. Consequently, monitoring RH is critical for environmental and public health planning. Both low and high RH levels pose risks: low RH is associated with fires, water scarcity, and health complications (Toxicology, Aircraft, Studies, \& Council, 2002), while high RH can trigger respiratory, cardiac, and rheumatic issues-highlighting the importance of monitoring extreme percentiles rather than averages in RH data.\\
\hspace*{0.2in} The RH data used in this application was accessed from the Haarweg Wageningen weather station of Wageningen University (Netherlands, 2009) for the month of May 2007 and 2008 and was reported by Raschke (2011) as follows.\\
May 2007: 0.4, 0.44, 0.5, 0.55, 0.58, 0.62, 0.65, 0.69,
0.72, 0.72, 0.73, 0.75, 0.77, 0.8, 0.81, 0.81, 0.83, 0.83, 0.85,
0.85, 0.85, 0.85, 0.86, 0.86, 0.87, 0.87, 0.89, 0.92, 0.94,
0.94, 0.97\\

May 2008: 0.39, 0.4, 0.42, 0.43, 0.43, 0.43, 0.44, 0.46,
0.48, 0.49, 0.51, 0.52, 0.53, 0.54, 0.56, 0.59, 0.62, 0.64,
0.66, 0.73, 0.75, 0.76, 0.83, 0.85, 0.88, 0.91, 0.92, 0.92,
0.95, 0.97, 0.98\\
The summary measures of the RH data for May 2007 and 2008 are given below.    
\begin{center}
\[
\begin{array} {ccccccccc}
2007   &  &  &  & &  &  &  & \\
 Min     & \hspace*{0.1 in}5\%  &\hspace*{0.1 in}  10\%    & \hspace*{0.1 in}  25\%   &\hspace*{0.1 in}  50\%  &\hspace*{0.1 in}    75\%  &\hspace*{0.1 in}  90\%  &\hspace*{0.1 in}   95\%   &\hspace*{0.1 in}  Max\\
0.440  & \hspace*{0.1 in}0.522 &\hspace*{0.1 in} 0.577   &\hspace*{0.1 in} 0.720  &\hspace*{0.1 in}   0.820  &\hspace*{0.1 in}   0.860 &\hspace*{0.1 in}  0.922  &\hspace*{0.1 in}  0.940  &\hspace*{0.1 in} 0.970\\
	2008   &  &  &  & &  &  &  & \\
 		Min     & \hspace*{0.1 in}5\%  &\hspace*{0.1 in}  10\%    & \hspace*{0.1 in}  25\%   &\hspace*{0.1 in}  50\%  &\hspace*{0.1 in}    75\%  &\hspace*{0.1 in}  90\%  &\hspace*{0.1 in}   95\%   &\hspace*{0.1 in}  Max\\
 		 		0.400  & \hspace*{0.1 in}0.424 &\hspace*{0.1 in} 0.430   &\hspace*{0.1 in} 0.482  &\hspace*{0.1 in}   0.605  &\hspace*{0.1 in}   0.845 &\hspace*{0.1 in}  0.923  &\hspace*{0.1 in}  0.961  &\hspace*{0.1 in} 0.980
 	\end{array}
 	\]
 \end{center}
 
\begin{figure}[H]
	\centering
	\begin{minipage}[b]{0.48\linewidth}
		\includegraphics[height=5.5 cm,width=8 cm]{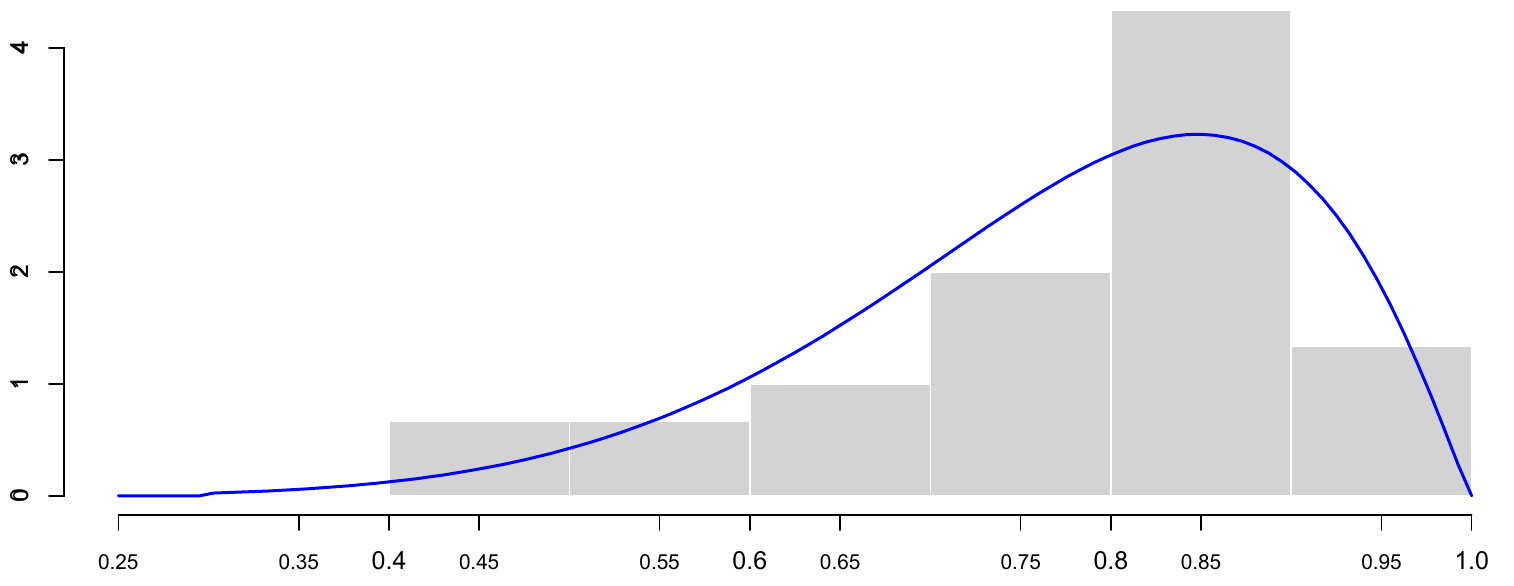}
		\centering{$\left(a\right)$ Histogram and density plot for the data of May 2007}
	\end{minipage}
	\quad
	\begin{minipage}[b]{0.48\linewidth}
		\includegraphics[height=5.5 cm,width=8 cm]{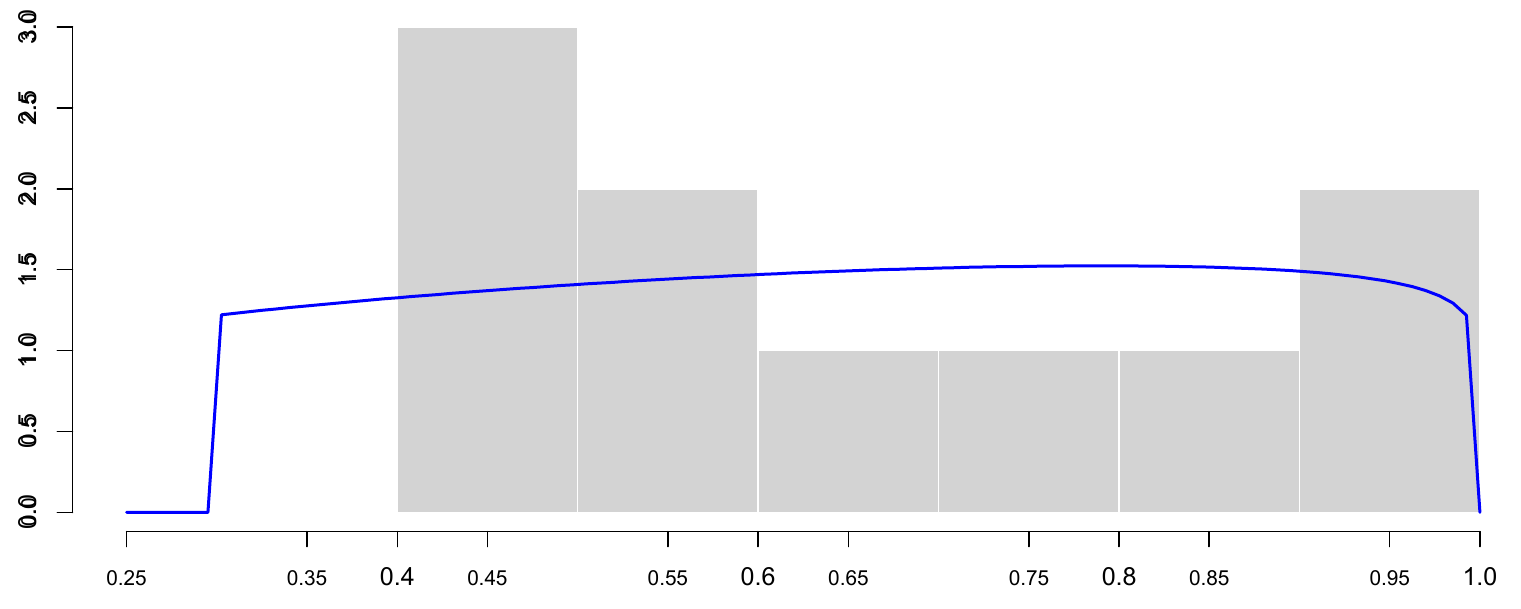}
		\centering{$\left(b\right)$ Histogram and density plot for the data of May 2008}
	\end{minipage}\caption{Histogram and density plot for relative humidity data}\label{figure1}
\end{figure}

The histogram in Figure \ref{figure1} confirms the asymmetry nature of the RH data set. 

\begin{figure}[H]\centering
	\includegraphics[height=5.5 cm,width=15 cm]{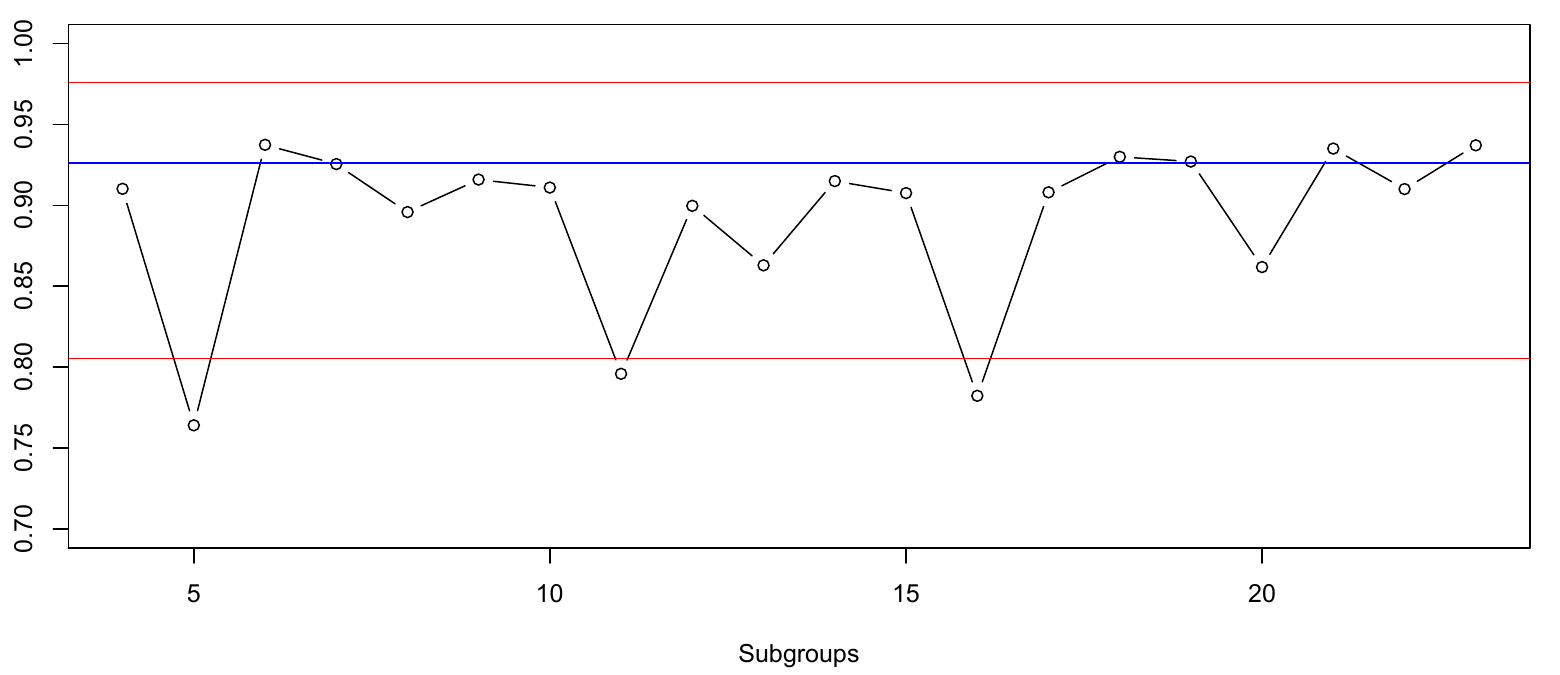}
	\caption{ $SBTBC$ Control chart for $90^{th}$ percentile of RH data data based on $20$ OOC subgroup with $FAR=0.0027$  and $\Delta\theta_1=-0.2$, $\Delta\theta_2=0$, $UCL=0.976$, $CL= 0.926$, $LCL=0.805.$, $a=0.3$, $b=1$. }\label{figure12}
\end{figure}

Excluding the first observation, the remaining thirty observations from the RH data for May 2007 are divided into three in-control (IC) samples of size \( n = 10 \) each (i.e., \( k = 3 \)). Based on these samples, MLEs of $\theta_1$ and $\theta_2$ of the Tbeta distribution with support $(0.3, 1)$ are obtained as $7.448$ and $2.154$ respectively with $\xi_{0.90}=0.922$ as the $90^{th}$ percentile. The Kolmogorov–Smirnov (K-S) test yields a statistic of $KS=0.1138$ and a $p$-value of 0.8317, indicating that the Tbeta distribution adequately fits the data. For a bootstrap sample size $n=10$, bootstrap size $B=5,000,$ $FAR=0.0027$, MLEs of $\theta_1$ and $\theta_2,$ and following the steps 3-7 in Subsection 2.2, the control limits for the $90^{th}$ percentile of $SBTBC$ chart are determined as $UCL=0.976
$, $CL=0.926$, $LCL=0.805.$ After establishing the three in-control subgroups, $20$ OOC subgroups of size $n=10$ each are generated from an OOC process with a $20\%$ decrease in $\theta_1,$ while keeping $\theta_2$ unchanged. The OOC performance of the $90^{th}$ percentile $SBTBC$ chart for this shift is presented in Figure $\ref{figure12},$ which shows that the chart's strong ability to detect the OOC signals, with $3$ of the $90^{th}$ percentile points falling below the $LCL.$ \\

\begin{figure}[H]\centering
	\includegraphics[height=5.5 cm,width=15 cm]{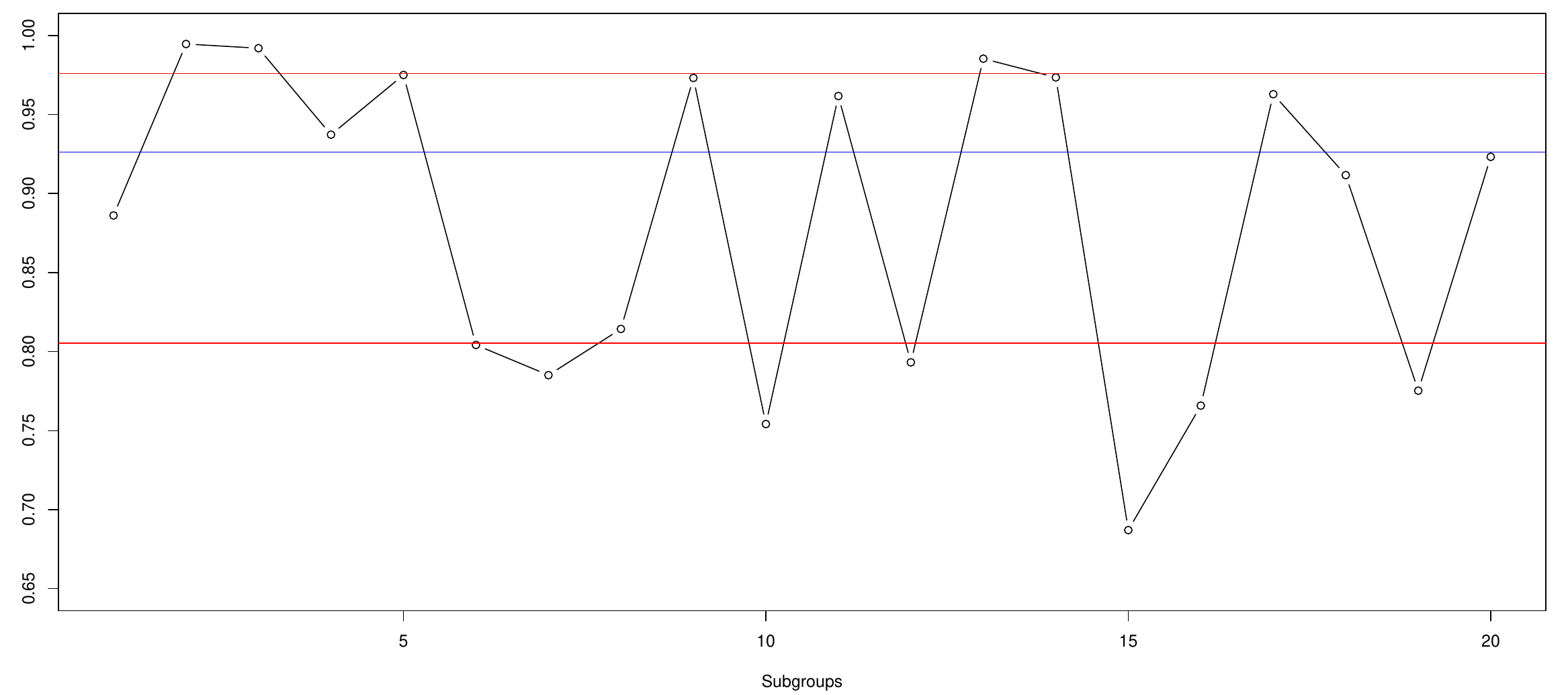}
	\caption{ $SBTBC$ Control chart for $90^{th}$ percentile using RH data of May 2008 on $20$ subgroup with $FAR=0.0027$ and $UCL=0.976$, $CL= 0.926$, $LCL=0.805.$, $a=0.3$, $b=1$}\label{figure18.1}
\end{figure}

\hspace*{0.2in} Now, for the May 2008 data, there are 31 observations. Excluding the first, the remaining $30$ observations are divided into three subgroups of size $n=10$ each. Based on these samples, the MLEs of the parameters \( \theta_1 \) and \( \theta_2 \) of the Tbeta distribution with support \( (0.3, 1) \) are obtained as \( \hat{\theta}_1 = 1.344 \) and \( \hat{\theta}_2 = 1.091 \), respectively. These values represent a substantial decrease compared to the May 2007 data-approximately 82\% reduction in \( \theta_1 \) and a 49\% reduction in \( \theta_2 \). The corresponding 90\textsuperscript{th} percentile is estimated as \( \xi_{0.90} = 0.926 \), slightly higher than the value from May 2007. The goodness-of-fit of the Tbeta distribution is assessed using the $K-S$ test, which gives a test statistic of $K-S = 0.127$ and a \( p \)-value of 0.714. These results suggest that the Tbeta distribution fits the May 2008 data well. Using the obtained MLEs, a data set of 30 observations is generated from the fitted Tbeta distribution with support \( (0.3, 1) \). Subsequently, 20 bootstrap samples of size \( n = 10 \) are drawn from this data set, and the corresponding 90\textsuperscript{th} percentiles are computed for each sample. These percentiles are then plotted against the control limits obtained from the May 2007 data. The resulting control chart, shown in Figure~\ref{figure18.1}, reveals that approximately $11$ out of $20$  percentiles fall outside the control limits, indicating that the RH data from May 2008 is out of control.

\begin{figure}[H]\centering
	\includegraphics[height=5.5 cm,width=15 cm]{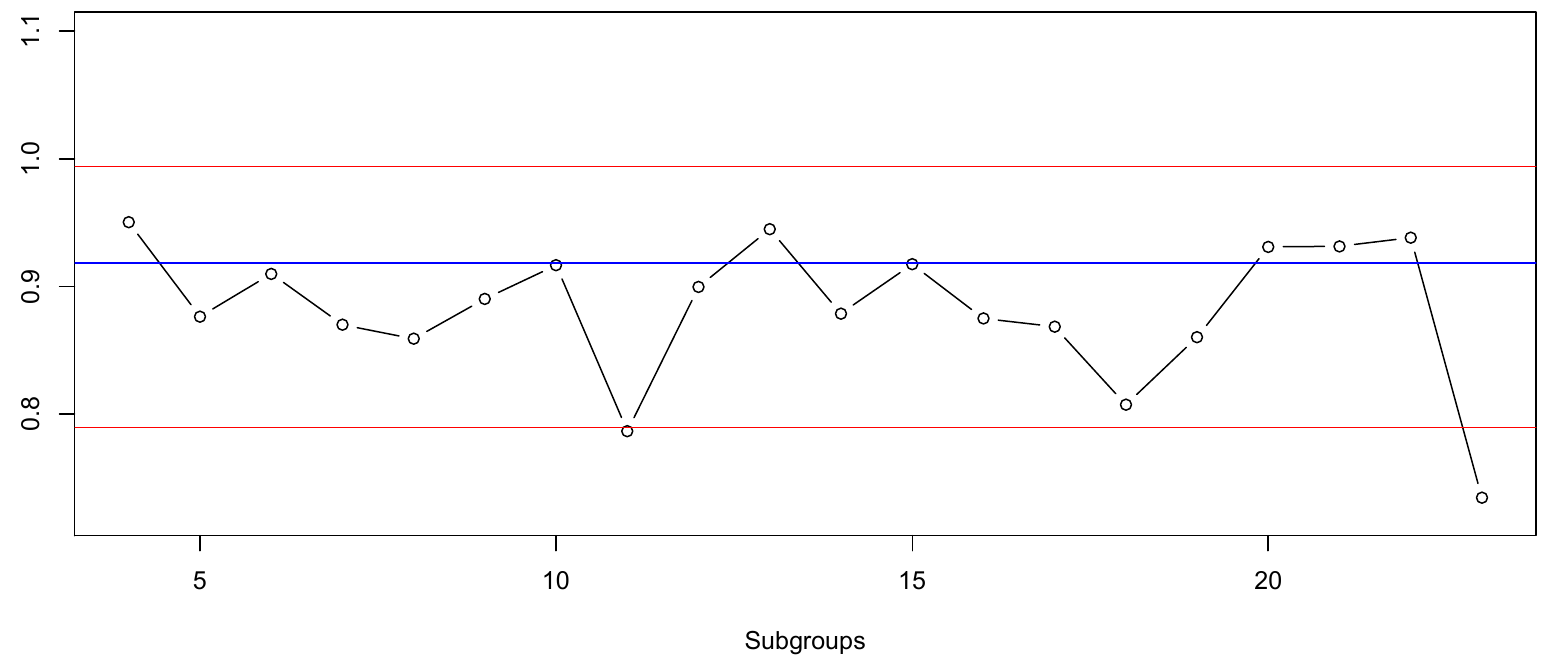}
	\caption{ $SBBC$ Control chart for $90^{th}$ percentile of RH data based on $20$ OOC subgroup with $FAR=0.0027$  and $\Delta\theta_1=-0.2$, $\Delta\theta_2=0$, $UCL= 0.994$, $CL= 0.918$, $LCL=  0.789.$ }\label{figure13}
\end{figure}

\hspace*{0.2in} The $SBTBC$ chart is now compared with the studentized bootstrap beta control chart ($SBBC$ chart). Beta distribution is found to fit the RH data equally well. The MLEs of $\theta_1$ and $\theta_2$ of beta distribution are obtained as $7.535$ and $2.171$ respectively. Following similar Steps $3-7$ in Subsection $2.2.$, control limits of $SBBC$ chart for the $90^{th}$ percentile are found to be $UCL=0.994$, $CL=0.918$, $LCL=0.789$. The chart statistics for twenty OOC subgroups of size $n=10$ each, generated from an OOC beta process having $\Delta\theta_1=-0.2, \Delta\theta_2  =0$ are plotted in Figure $\ref{figure13}$. Comparing both the charts, it is found that the $SBTBC$ chart has out-performed the $SBBC$ chart in terms of both frequency and speed. While $SBTBC$ chart has been able to generate the first OOC signal at the $2^{nd}$ test sample, the same for $SBBC$ chart is obtained at sample number $7$. Moreover, the number of OOC signals for $SBTBC$ chart is significantly higher than that of the $SBBC$ chart. This further justifies the use of $SBTBC$ chart for percentiles of proportion data over $SBBC$ chart.\\

\subsection{Monitoring Flood Levels}
\setcounter{equation}{0}
\hspace*{0.2in} This section illustrates another application of the $SBTBC$ chart to a hydrological data, with a focus on monitoring extreme flood levels. The control chart is designed to detect increase in the upper percentile of flood levels, which is of critical importance in flood risk management and infrastructure planning. In particular, higher percentiles provide early warnings of potential extreme levels that may indicate significant risks to life and property.

The data, originally compiled by Dumonceaux and Antle (1973), consist of the maximum flood levels (in millions of cubic feet per second) for the Susquehanna River at Harrisburg, Pennsylvania. Each observation represents the highest flood level recorded over a four-year period, beginning with $0.654$ for $1890-1893.$ The complete data set comprises $20$ such aggregated observations spanning a long historical time frame. The summary of the data are given below:
\begin{center}
	\[
	\begin{array} {ccccccccc}
		Min     & \hspace*{0.1 in}5\%  &\hspace*{0.1 in}  10\%    & \hspace*{0.1 in}  25\%   &\hspace*{0.1 in}  50\%  &\hspace*{0.1 in}    75\%  &\hspace*{0.1 in}  90\%  &\hspace*{0.1 in}   95\%   &\hspace*{0.1 in}  Max\\
		
		0.260  & \hspace*{0.1 in}0.269 &\hspace*{0.1 in} 0.297   &\hspace*{0.1 in} 0.335  &\hspace*{0.1 in}   0.405  &\hspace*{0.1 in}   0.457&\hspace*{0.1 in}  0.614  &\hspace*{0.1 in}  0.654  &\hspace*{0.1 in} 0.740
	\end{array}
	\]
\end{center}

\begin{figure}[H]
	\centering
		\includegraphics[height=5.5 cm,width=8 cm]{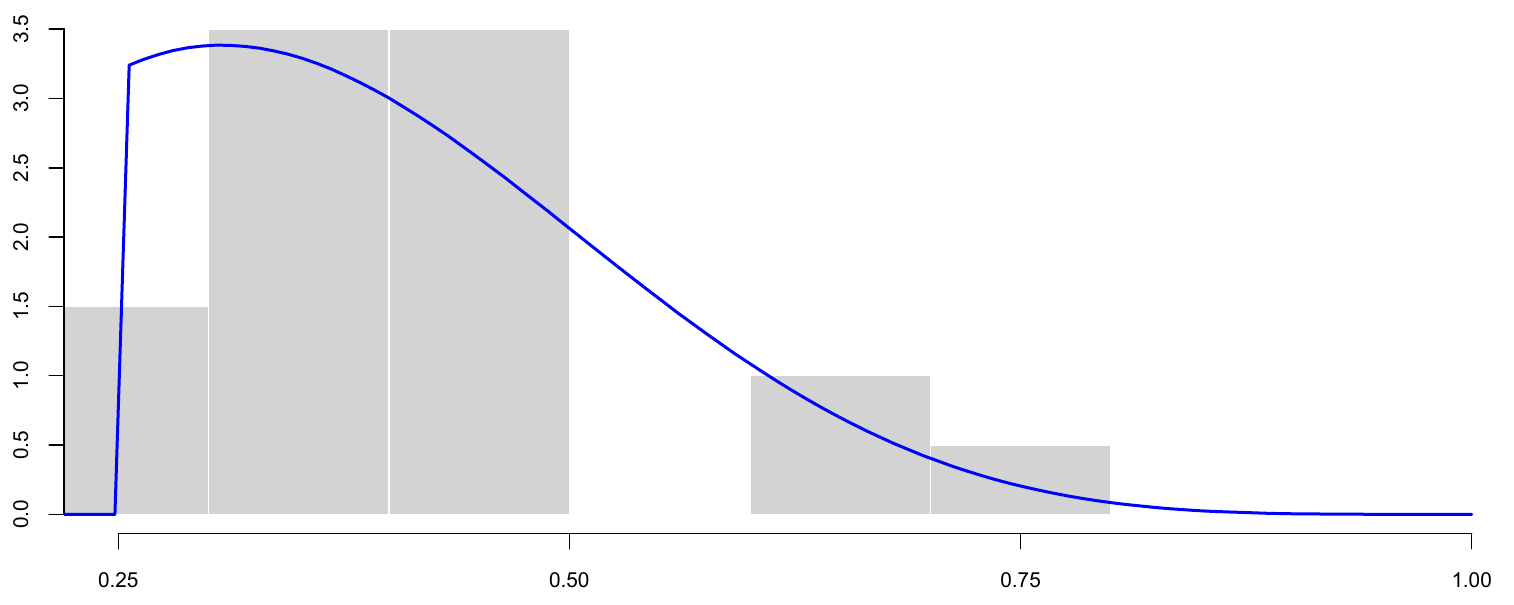}
		\caption{Histogram and density plot for maximum flood level data}\label{fig24}
	\end{figure}

\begin{figure}[H]\centering
	\includegraphics[height=5.5 cm,width=15 cm]{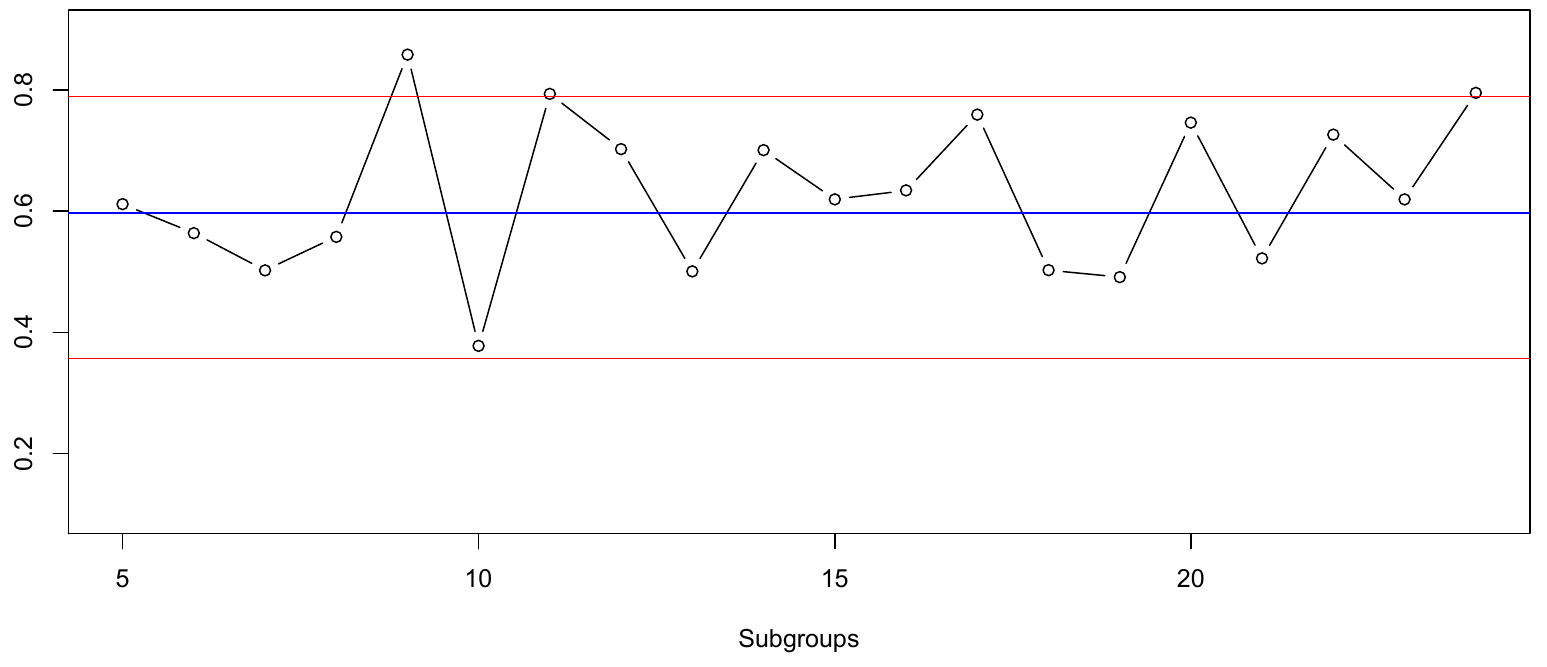}
	\caption{ $SBTBC$ Control chart for $90^{th}$ percentile of maximum flood level data  based on $20$ OOC subgroup with $FAR=0.0027$  and $\Delta\theta_1=0.2$, $\Delta\theta_2=0$, $UCL= 0.789$, $CL= 0.597$, $LCL=  0.357.$ }\label{figure14}
\end{figure}

\begin{figure}[ht]
	\centering
	\begin{minipage}[b]{0.48\linewidth}
		\includegraphics[height=5.2 cm, width=7.2 cm]{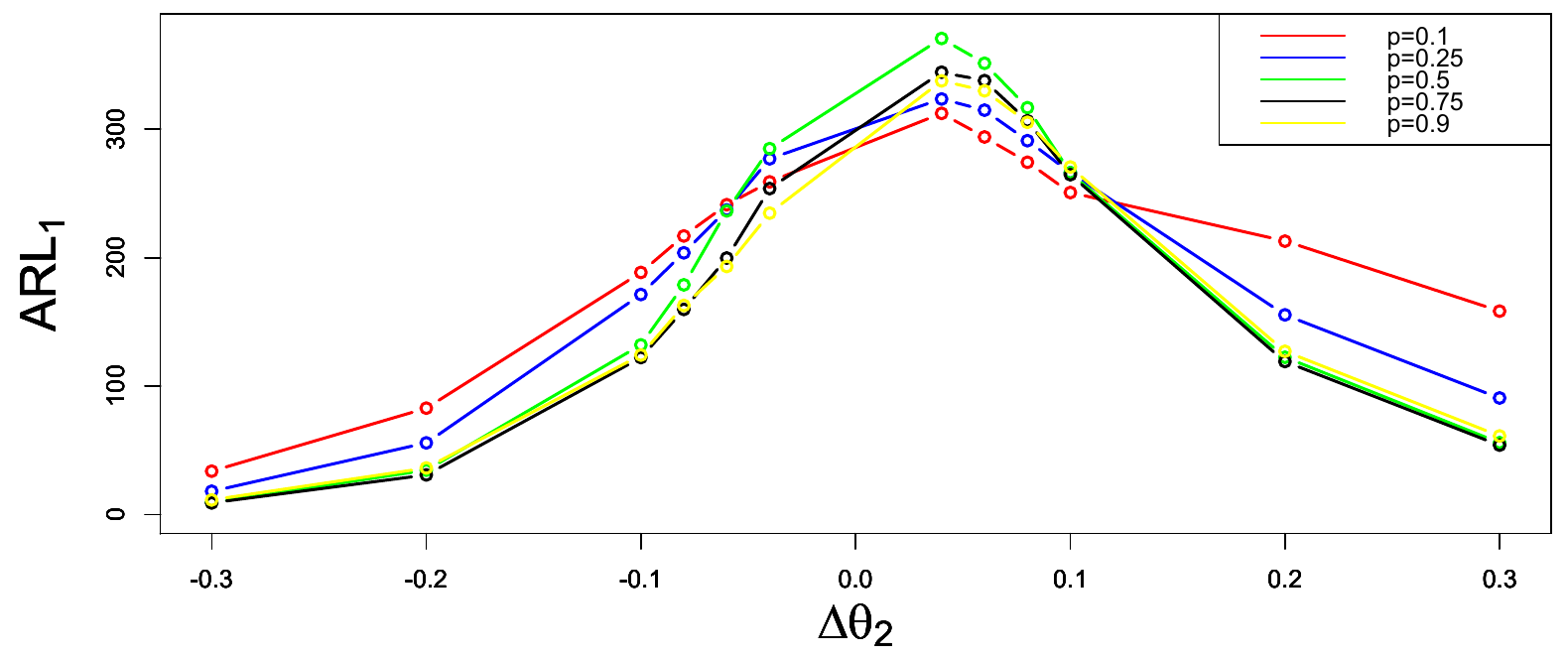}
		\centering{$\left(a\right)$ $ARL_1$ for different choices of $\Delta\theta_2$, when $\Delta\theta_1=0$}
	\end{minipage}
	\quad
	\begin{minipage}[b]{0.48\linewidth}
		\includegraphics[height=5.2 cm, width=7.2 cm]{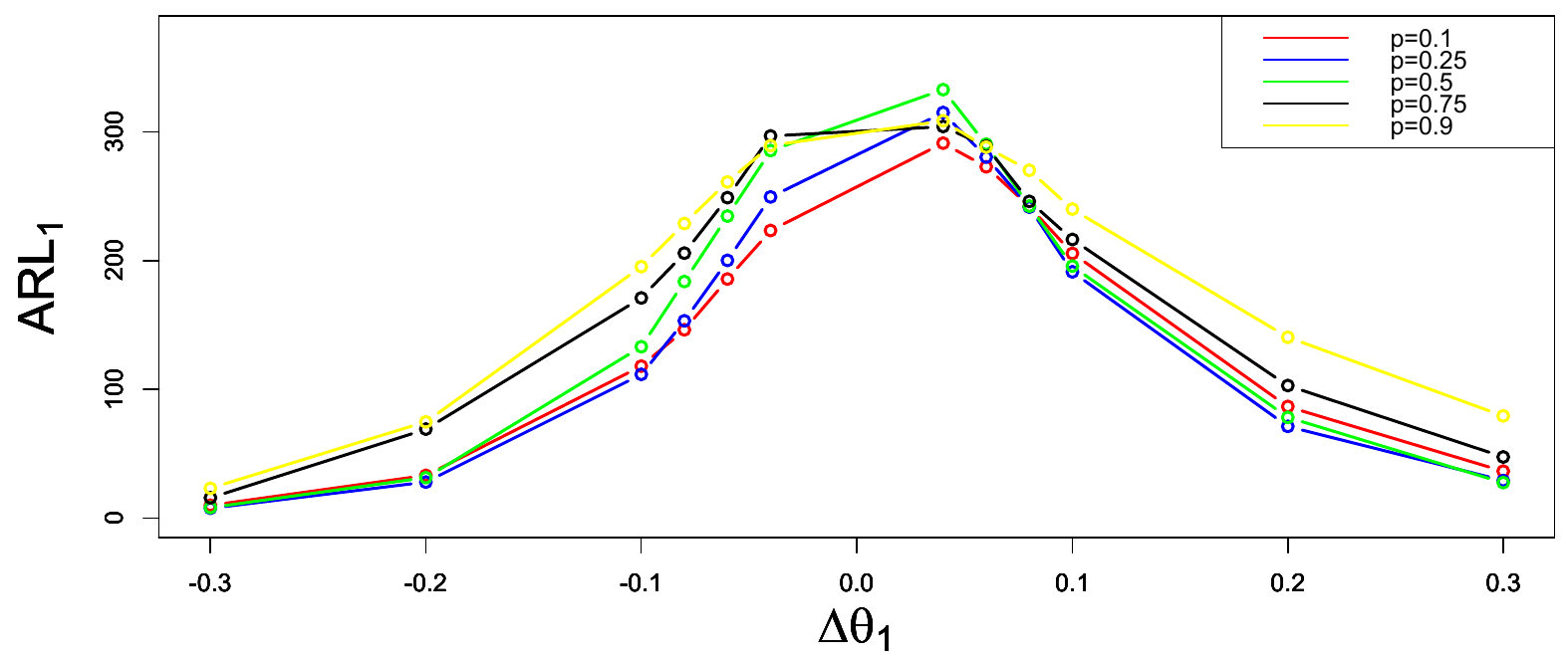}
		\centering{$\left(b\right)$ $ARL_1$ for different choices of $\Delta\theta_1$, when $\Delta\theta_2=0$}
	\end{minipage}
	\quad
	\begin{minipage}[b]{0.48\linewidth}
		\includegraphics[height=5.2 cm, width=7.2 cm]{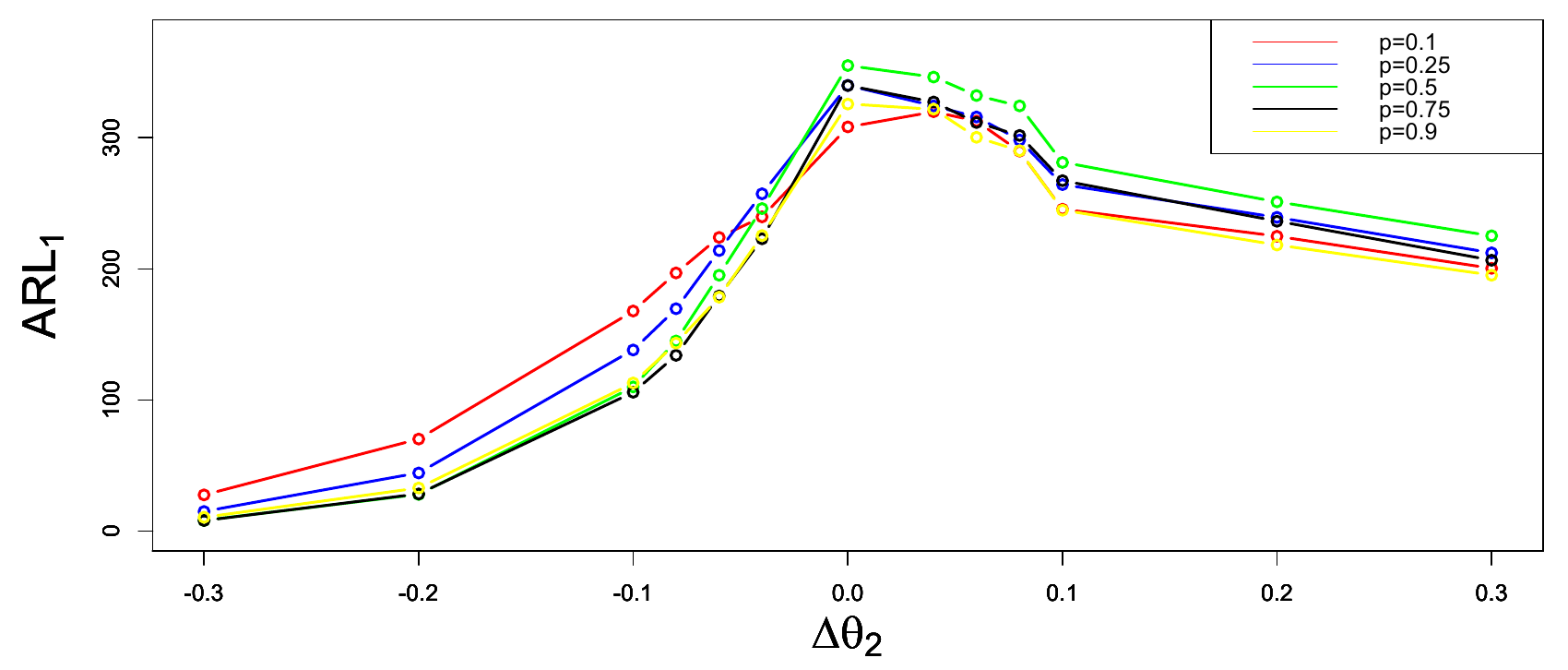}
		\centering{$\left(c\right)$ $ARL_1$ for different choices of $\Delta\theta_2$, when $\Delta\theta_1=0.02$}
	\end{minipage}
	\quad
	\begin{minipage}[b]{0.48\linewidth}
		\includegraphics[height=5.2 cm, width=7.2 cm]{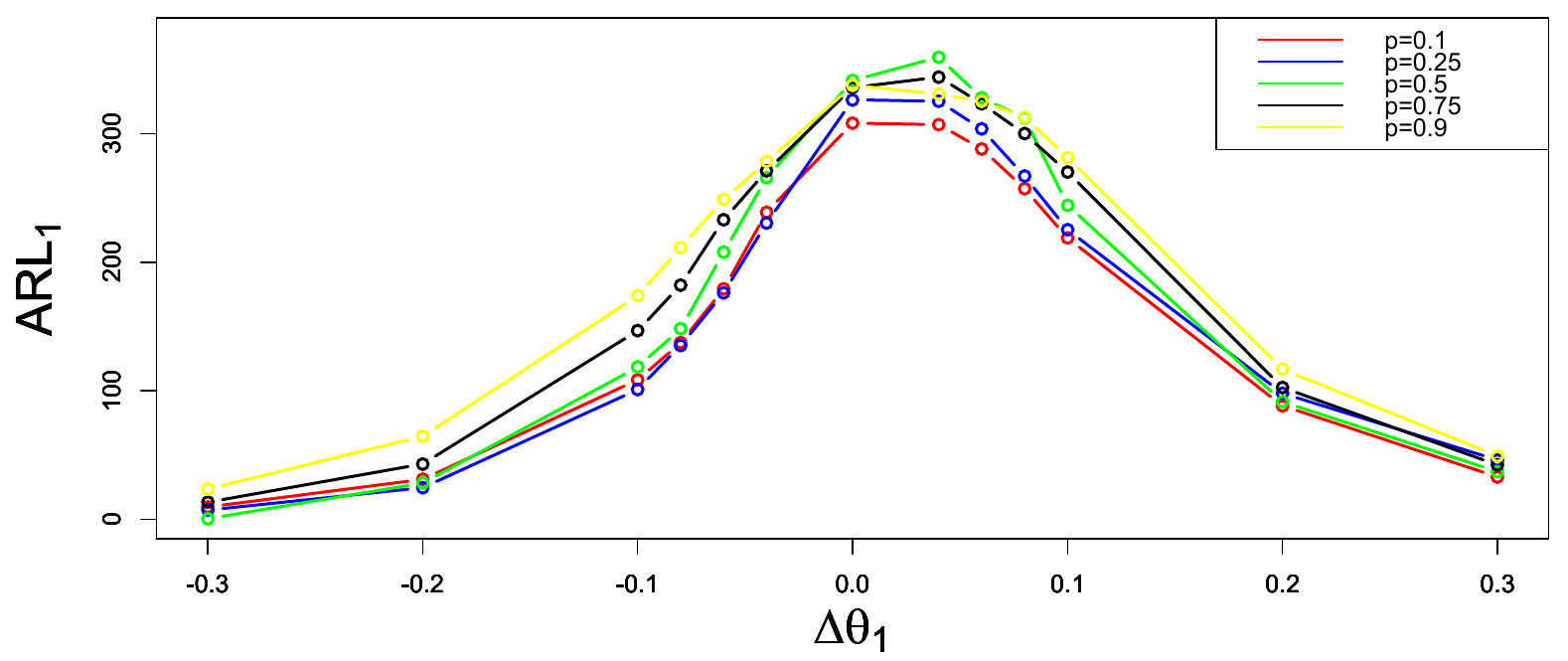}
		\centering{$\left(d\right)$ $ARL_1$ for different choices of $\Delta\theta_1$, when $\Delta\theta_2=0.02$}
	\end{minipage}\caption{\label{OOC} Graphs of $ARL_1$ for $SBTBC\theta$ schemes for $\nu=0.0027$.}
\end{figure}

\begin{figure}[H]
\centering
\begin{minipage}[b]{1\linewidth}
\includegraphics[height=6.8 cm,width=16 cm]{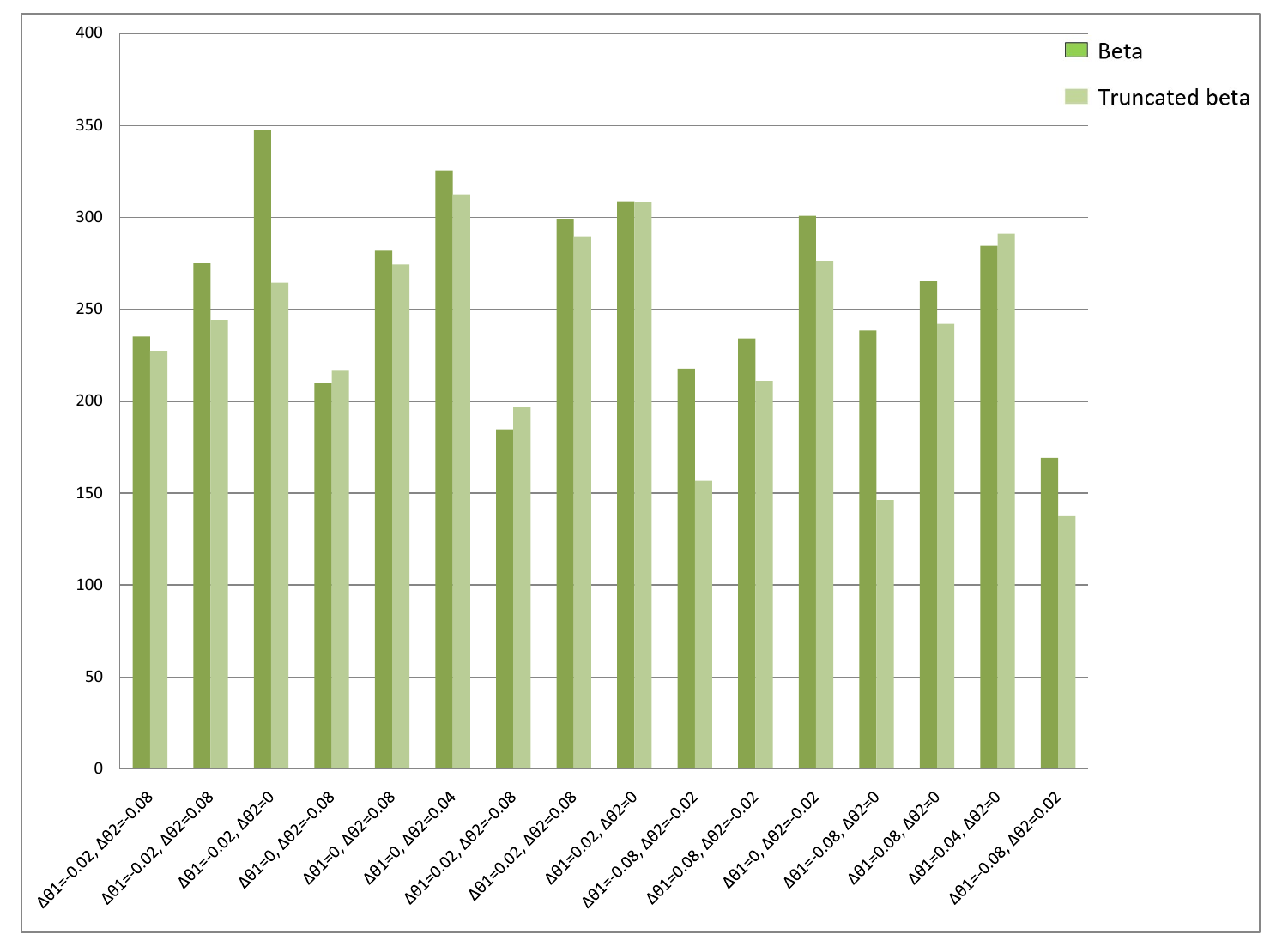}
\caption{Comparison of $ARL_1$ for different choices of $\Delta\theta_1$ and $\Delta\theta_2$ when $p=0.1$}\label{figure15}
\end{minipage}
\end{figure}

\begin{figure}[ht]
\centering
\includegraphics[height=6.8 cm,width=16 cm]{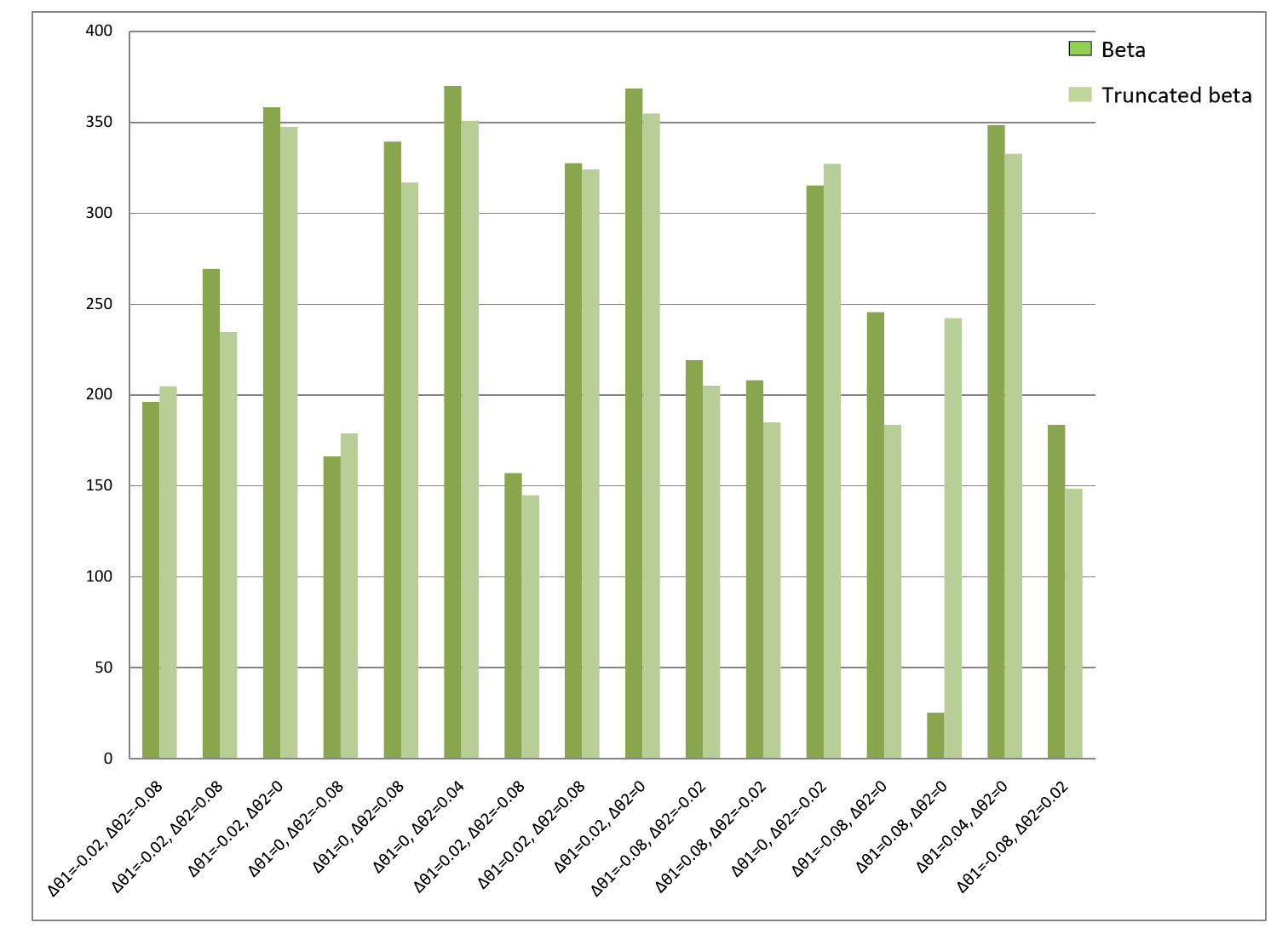}
\caption{Comparison of $ARL_1$ for different choices of $\Delta\theta_1$ and $\Delta\theta_2$ when $p=0.5$}\label{figure16}
\end{figure}

\begin{figure}[ht]
\centering
\includegraphics[height=6.8 cm,width=16 cm]{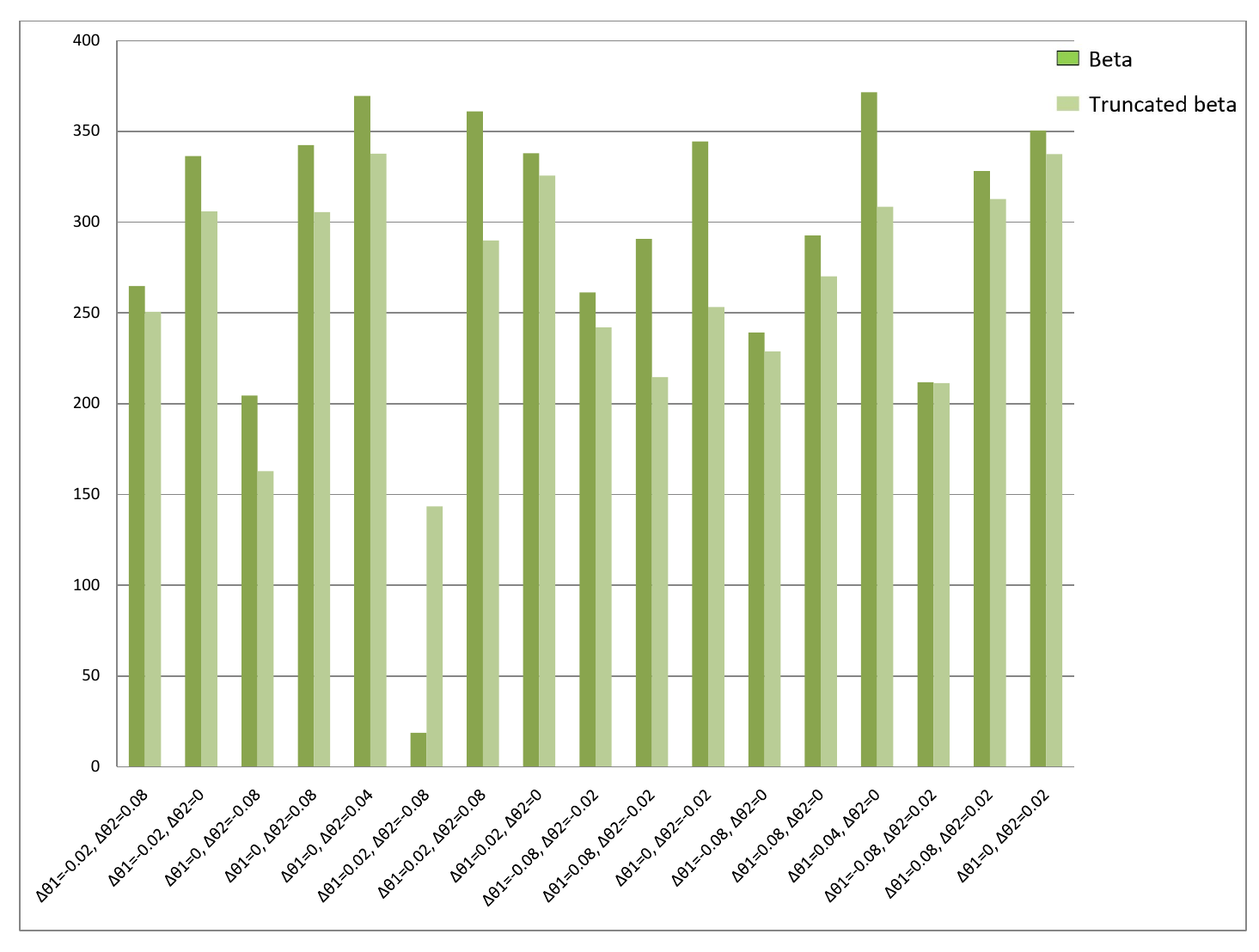}
\caption{Comparison of $ARL_1$ for different choices of $\Delta\theta_1$ and $\Delta\theta_2$ when $p=0.9$}\label{figure17}
\end{figure}

There are twenty observations split into four $(k=4)$ IC samples of size $n=5$ each basis which the MLEs of $\theta_1$ and $\theta_2$ of the Tbeta distribution with support $(0.25, 1)$ are obtained as $3.003$ and $5.510$ respectively. The fitted density is provided in Figure~\ref{fig24}. The $K-S$ statistic value $(KS=0.149)$ and the $p$-value (=0.762) suggest that the Tbeta distribution fits the data well. For bootstrap sample size $n=5$, bootstrap size $B=5000,$ $FAR=0.0027$, MLEs of $\theta_1$ and $\theta_2,$ and following the steps 3-7 in Subsection 2.2, the control limits for the $90^{th}$ percentile $SBTBC$ chart are found to be $UCL=0.789
$, $CL=0.597$, $LCL=0.357$. After the first four in-control subgroups, $20$ OOC  subgroups of size $n=5$ each are generated from an OOC process that has different shape parameters with $20\%$ increase in $\theta_1$ and in-control $\theta_2.$ The OOC performance of the control chart for $90^{th}$ percentile for the shift mentioned is presented in Figure $\ref{figure14}$ which shows that the $SBTBC$ chart performs very well to detect the OOC signals with $3$ of the $90^{th}$ percentile points falling above the $UCL$.  \\

\setcounter{section}{4}
\section{Concluding Remark}
\setcounter{equation}{0}
Standard $p$ and $np$ charts, as well as their modified versions, may often be inadequate for reliably monitoring proportion-type data across a wide range of practical scenarios. The beta distribution, defined on the interval [0, 1], has long been a popular and flexible choice for modeling proportion data due to its diverse shapes and broad applicability. However, in many real-world contexts, processes do not allow for extremely high or low proportions of nonconforming items. In such cases, right or left truncation of the beta distribution can more accurately represent the inherent variability in the proportion data compared to the un-truncated beta distribution. Building on this rationale, this paper introduces a control chart for monitoring percentiles of a process that follows a truncated beta distribution. Studentized parametric bootstrap method is used to estimate the in-control parameters and the control limits. The IC and OOC performances of the proposed chart are evaluated through average run length in a comprehensive simulation study. The simulated IC $ARL$ values closely align with the theoretical results, demonstrating the effectiveness of the chart for skewed data. Moreover, the OOC $ARL$ values decrease sharply in response to small, medium and large shifts in the parameters-both downward and upward, confirming the chart's sensitivity and speed in detecting shifts. The performance of the proposed chart is validated using environmental and climate data sets, specifically, relative humidity and flood levels data. Results also demonstrate that the proposed chart outperforms the conventional beta-based control chart. 

{\bf Disclosure Statement}\\
No potential conflict of interest was reported by the author(s).\\
{\bf Acknowledgment}\\ 
Research scholarship grant from NBHM (No. 0203/4/2020/R\&D-II/2962 dated 22.02.2024) is acknowledged with thanks by Bidhan Modok.

\newpage



\begin{thebibliography}{99}
\bibitem {ba1} Bayer, F. M., Tondolo, C. M.and Müller, F. M. (2018), ``Beta regression control chart for monitoring fractions and proportions",{\it Computers and Industrial Engineering}, Vol. 119, pp. 416-426.
\bibitem {ba2} Bayer, F. M. and Cribari-Neto, F. (2017), ``Model selection criteria in beta regression with varying dispersion" {\it  Communications in Statistics - Simulation and Computation}, Vol. 46 No. 1, pp. 729-746.
\bibitem {be} Bersimis, S., Koutras, M. V. and Maravelakis, P. E. (2014), ``A compound control chart for
monitoring and controlling high quality processes", {\it European Journal of Operational Research}, Vol. 233 No. 3, pp. 595-603.
\bibitem{bi} Biktasheva, I. V. (2020). Role of a habitat's air humidity in Covid-19 mortality. {\it Science of the Total Environment}, 736, 138763.
\bibitem{ca} Camuffo, D. (1998). Chapter 2 humidity. {\it Microclimate for Cultural Heritage}, 23, 42–89.
\bibitem {che} Chen, G. (1998), ``An improved p chart through simple adjustment", {\it Journal of Quality Technology}, Vol. 30, pp. 142-151.
\bibitem {chi1} Chiang, J. Y., Jiang, N., Brown, T. N., Tsai, T. R. and Lio, Y. L. (2017), ``Control charts for generalized exponential distribution percentiles", {\it Communications in Statistics-Simulation and Computation}, Vol. 46 No. 10, pp. 7827-7843.
\bibitem {chi2} Chiang, J. Y., Lio, Y. L., Ng, H. K. T., Tsai, T. R. and Li, T. (2018), ``Robust bootstrap control charts for percentiles based on model selection approaches", {\it Computers \& Industrial Engineering}, Vol. 123, pp. 119-133.
\bibitem {cho1} Chowdhury, S., Kundu, A., \& Modok, B. (2022). ``Bootstrap beta control chart for monitoring proportion data", {\it International Journal of Quality \& Reliability Management}, Vol. 39(10), pp. 2354-2377.
\bibitem {cho2} Chowdhury, S., Kundu, A., and Modok, B. (2025). Control Monitoring Schemes for Monitoring Percentiles of Generalized Exponential Distribution with Hybrid Censoring. {\it REVSTAT-Statistical Journal}, 23(1), 1-17.
\bibitem{dea1} de Araujo Lima-Filho, L. M., and Mariano Bayer, F. (2021). Kumaraswamy control chart for monitoring double bounded environmental data. {\it Communications in Statistics-Simulation and Computation}, 50(9), 2513-2528.
\bibitem {dea} de Araujo Lima-Filho, L. M., Pereira, T. L., de Souza, T. C. and Bayer, F. M. (2019), ``Inflated beta control chart for monitoring double bounded processes", {\it Computers \& Industrial Engineering}, Vol. 136, pp. 265-276.
\bibitem {dr} Draper, N. R. and Smith, H. (1998), {\it Applied regression analysis (Vol. 326)}, John Wiley \& Sons, New York, NY.
\bibitem{du} Duclos, A. and Voirin, N. (2010), ``The p-control chart: a tool for care improvement", {\it International Journal for Quality in Health Care}, Vol. 22(5), pp. 402-407.
\bibitem{duc}Dumonceaux, R. and Antle, C. E. (1973), ``Discrimination between the log-normal and the Weibull distributions", {\it Technometrics},  Vol. 15(4), pp.923-926.
\bibitem{ef1} Efron, B. (1979). ``Computers and the theory of statistics: thinking the unthinkable", {\it SIAM review}, Vol. 21(4), pp. 460-480.
\bibitem {ef} Efron, B. and Tibshirani R. J (1993), {\it An Introduction to the Bootstrap}, Chapman \& Hall, New York, NY.
\bibitem{ga} Gao, J., Sun, Y., Lu, Y., and Li, L. (2014). Impact of ambient humidity on child health: a systematic review. {\it PloS one}, 9(12), e112508.
\bibitem {gu1} Gupta, R. D. and Kundu, D. (2001), ``Generalized exponential distribution: different method of estimations", {\it Journal of Statistical Computation and Simulation}, Vol. 69 No. 4, pp. 315-337.
\bibitem {gu} Gupta, A. K. and Nadarajah, S. (2004), {\it Handbook of beta distribution and its applications}, CRC Press LLC.
\bibitem {he} Heimann, P. A. (1996), ``Attribute control charts with large sample sizes", {\it Journal of Quality Technology}, Vol. 28 No. 4, pp. 451-459.
\bibitem{ho} Ho, L. L., Fernandes, F. H. and Bourguignon, M. (2019), ``Control charts to monitor rates and proportions", {\it Quality and Reliability Engineering International}, Vol. 35, pp. 74-83.
\bibitem {hy} Hyndman, R. J. and Fan, Y. (1996), ``Sample quantiles in statistical packages", {\it American Statistician}, Vol. 50, pp. 361-365.
\bibitem {je} Jensen, W. A., Jones-Farmer, L. A., Champ, C. W. and Woodall, W. H. (2006), ``Effects of parameter estimation on control chart properties: a literature review", {\it Journal of Quality Technology}, Vol. 38 No. 4, pp. 349-364.
\bibitem {jo} Johnson, N., Kotz, S. and Balakrishnan, N. (1995), {\it Continuous univariate distributions}, Wiley series in probability and mathematical statistics: Applied probability and statistics, Vol. 2., Wiley \& Sons, New York, NY.
\bibitem {jon} Jones, L. A. and Woodall, W. Y. (1998), ``The performance of bootstrap control charts", {\it Journal of Quality Technology}, Vol. 30, pp. 362-375.
\bibitem{lim} Lima–Filho, L. M. A., Pereira, T. L., Bayer, F. M., de Souza, T. C., and Bourguignon, M. (2023). Control chart for monitoring zero-or-one inflated double-bounded environmental processes. {\it Environmental and Ecological Statistics}, 30(3), 355-377.
\bibitem {li} Lio, Y. L. and Park, C. (2008), ``A bootstrap control chart for Birnbaum-Saunders percentiles", {\it Quality and Reliability Engineering International}, Vol. 24, pp. 585-600. 
\bibitem {li1} Lio, Y. L. and Park, C. (2010), ``A bootstrap control chart for inverse Gaussian percentiles", {\it Journal of Statistical Computation and Simulation}, Vol. 80, pp. 287-299.
\bibitem {li2} Lio, Y.L., Tsai, T.-R., Aslam, M. and Jiang, N. (2014), ``Control charts for monitoring Burr type-X percentiles", {\it Communications in Statistics-Simulation and Computation}, Vol. 43, pp. 761-776.
\bibitem {li3} Liu, R. Y. and Tang, J. (1996), ``Control charts for dependent and independent measurements based on the bootstrap", {\it Journal of the American Statistical Association}, Vol. 91, pp. 1694-1700.
\bibitem{lu} Liu, J., Zhou, J., Yao, J., Zhang, X., Li, L., Xu, X., ... and Zhang, K. (2020). Impact of meteorological factors on the COVID-19 transmission: A multi-city study in China. {\it Science of the total environment}, 726, 138513.
\bibitem{mod} Modok, B., Chowdhury, S., and Kundu, A. (2025). Process monitoring schemes for generalized Weibull quantiles under hybrid censoring. {\it Quality Technology \& Quantitative Management}, 1-24.
\bibitem{mod2} Modok, B., Chowdhury, S., and Kundu, A. (2025). Control charts for monitoring Weibull quantile under generalized hybrid and progressive censoring schemes. {\it Statistics and Computing}, 35(2), 49.
\bibitem {mo} Montgomery, D. C. (2009), {\it Introduction to statistical quality control (6th ed.)}, John Wiley \& Sons, New York, NY.
\bibitem {ni} Nichols M. D. and Padgett W. J. (2005), ``A bootstrap control chart for Weibull percentiles", {\it Quality and Reliability Engineering International}, Vol. 22, pp. 141-151.
\bibitem {pa} Papadakis, M., Tsagris, M., Dimitriadis, M., Fafalios, S., Tsamardinos, I., Fasiolo, M., Borboudakis, G., Burkardt, J., Zou, C., Lakiotaki, K. and Chatzipantsiou, C. (2020), ``A collection of efficient and extremely fast R functions, version 1.9.9. R Package version 4.0.0". URL: https://github.com/RfastOfficial/Rfast.
\bibitem {qu} Quesenberry, C.P. (1991), ``SPC Q charts for binomial parameter p: Short or long runs", {\it Journal of Quality Technology}, Vol. 23, pp. 239-246.
\bibitem{Ra} Raschke, M. (2011), ``Empirical behaviour of tests for the beta distribution and their application in environmental research", {\it Stochastic Environmental Research and Risk Assessment},  Vol. 25, No.1 pp.79-89.
\bibitem {sa} Sant’Anna, $\hat{\text A}$. M. O. and ten Caten, C. S. (2012), ``Beta control charts for monitoring fraction data", {\it Expert Systems with Applications}, Vol. 39 No. 11, pp. 10236-10243.
\bibitem{sa1} Sarkodie, S. A., and Owusu, P. A. (2020). Impact of meteorological factors on COVID-19 pandemic: Evidence from top 20 countries with confirmed cases. {\it Environmental Research}, 191, 110101.
\bibitem {sc} Schwertman, N. C. and Ryan, T. P. (1997), ``Optimal limits for attributes control charts", {\it Journal of Quality Technology}, Vol. 29 No. 1, pp. 86-98.
\bibitem {se} Seppala, T., Moskowitz, H., Plante, R. and Tang, J (1995 ), ``Statistical process control via the subgroup bootstrap", {\it Journal of Quality Technology}, Vol. 27, pp. 139-153. 
\bibitem {sh} Shewhart, W. A. (1924), Economic control of quality manufactured product. Princeton: Van Nostrand Reinhold, p. 170.
\bibitem{si} Silva, C. R. D., Lima Filho, L. M. D. A., Pereira, T. L., and Duarte Neto, P. J. (2025). Shewhart-type control chart based on unit gamma distribution inflated at zero or one. {\it Journal of Statistical Computation and Simulation}, 1-22.
\bibitem{to} Toxicology, B., Aircraft, C., Studies, D., and Council, N. (2002). The airliner cabin environment and the health of passengers and crew. National Academies Press.
\bibitem {wa} Wang, H. (2009), ``Comparison of control charts for low defective rate", {\it Computational Statistics \& Data Analysis}, Vol. 53 	No. 12, pp. 4210-4220.
\bibitem{we1} Wageningen University, Netherlands, Meteorology and air quality section (2009, download). Weather station Haarweg Wageningen. Data: http://www.met.wau.nl/(2009). Accessed 5 October 2009.
\bibitem {we} Wetherill, G.B. and Brown, D.W. (1991), {\it Statistical Process Control—Theory and Practice}, Chapman \& Hall, London.
\bibitem {xi} Xie, M. and Goh, T.N. (1993), ``Improvement detection by control charts for high yield processes", {\it International Journal of Quality and Reliability Management}, Vol. 10, pp. 24-31.
\bibitem{ ya1} Yang, X. D., Li, H. L., and Cao, Y. E. (2021). Influence of meteorological factors on the COVID-19 transmission with season and geographic location. {\it International Journal of Environmental Research and Public Health}, 18(2), 484.
\bibitem{ya} Yao, M., Zhang, L., Ma, J., and Zhou, L. (2020). On airborne transmission and control of SARS-Cov-2. {\it Science of The Total Environment}, 731, 139178.
\end{thebibliography}
\end{document}